\documentclass[lettersize,journal]{IEEEtran}

\usepackage{subcaption}
\usepackage{tikz}
\usetikzlibrary{tikzmark, positioning, fit, shapes.misc}
\usetikzlibrary{decorations.pathreplacing, calc}
\usepackage{authblk}
\usepackage{xurl}
\usepackage{amsmath} 
\usepackage{amssymb}
\usepackage{mathtools}
\usepackage{algorithm}
\usepackage[noend]{algpseudocode}
\usepackage{interval}
\usepackage{eurosym}
\usepackage{scalerel}
\usepackage{tikz}
\usepackage{lipsum}
\usetikzlibrary{shapes,arrows}
\usepackage{bm,times}
\usepackage{verbatim}
\usepackage{amsfonts}
\usepackage{array}
\usepackage{textcomp}
\usepackage{stfloats}
\usepackage{url}
\usepackage{graphicx}
\usepackage{cite}
\DeclareMathAlphabet{\mathpzc}{T1}{pzc}{m}{it}
\DeclareMathOperator*{\E}{\mathbb{E}}

\DeclareMathOperator*{\argmin}{arg\,min}
\pgfdeclarelayer{background}
\pgfdeclarelayer{foreground}
\pgfsetlayers{background,main,foreground}
\tikzstyle{sensor}=[draw, fill=blue!20, text width=5em, 
    text centered, minimum height=2.5em]
\tikzstyle{pvmodel}=[draw, fill=red!20, text width=20em, 
    text centered, minimum height=5em]
\tikzstyle{ann} = [above, text width=5em]
\tikzstyle{naveqs} = [sensor, text width=6em, fill=red!20, 
    minimum height=12em, rounded corners]

\tikzset{virtual/.style={coordinate},
block/.style={draw, inner sep=3pt}}

\begin{document}

\title{
Optimizing Photovoltaic Panel Quantity for Water Distribution Networks
}

\author{Mirhan Ürkmez, \and Carsten Kallesøe, \and Jan Dimon Bendtsen, \and John Leth \thanks{Manuscript received 21 October 2023. This work is funded by Independent Research Fund
Denmark (DFF). (\textit {Corresponding author: Mirhan Ürkmez.)}}\thanks{Mirhan Ürkmez, Carsten Kallesøe, Jan Dimon Bendtsen, and  John Leth are with the Department of Electronic Sytems, Aalborg University, DK-9220 Aalborg,
 Denmark (e-mail: mu@es.aau.dk; csk@es.aau.dk; dimon@es.aau.dk; jjl@es.aau.dk).}}



\maketitle

\begin{abstract}
The paper introduces a procedure for determining an approximation of the optimal amount of photovoltaics (PVs) for powering water distribution networks (WDNs) through grid-connected PVs. The procedure aims to find the PV amount minimizing the total expected cost of the WDN over the lifespan of the PVs. The approach follows an iterative process, starting with an initial estimate of the PV quantity, and then calculating the total cost of WDN operation. To calculate the total cost of the WDN, we sample PV power profiles that represent the future production based on a probabilistic PV production model. Simulations are conducted assuming these sampled PV profiles power the WDN, and pump flow rates are determined using a control method designed for PV-powered WDNs. Following the simulations, the overall WDN cost is calculated. Since we lack access to derivative information, we employ the derivative-free Nelder-Mead method for iteratively adjusting the PV quantity to find an approximation of the optimal value. The procedure is applied for the WDN of Randers, a Danish town. By determining an approximation of the optimal quantity of PVs, we observe a 14.5\% decrease in WDN costs compared to the scenario without PV installations, assuming a 25 year lifespan for the PV panels.
\end{abstract}


\section{INTRODUCTION}
Water distribution networks (WDNs) are complex infrastructures that provide drinkable water to consumers in a specific area. Since electricity consumption on WDNs constitutes around  $7\%-8\%$ of the world's total energy consumption \cite{WSSEnergy}, numerous researchers have explored the problem of scheduling water pumps to minimize the operational costs of WDNs \cite{9765702,Kallese2017PlugandPlayMP}. One potential solution involves powering WDNs using renewable energy sources like Photovoltaic (PV) panels, which are easy to install and maintain. Specifically, grid-connected PVs can address the intermittent energy production issue of PV panels by allowing the grid to power WDN when there is insufficient PV power. To implement this solution,  one must determine the appropriate number of PV panels to install for a given WDN. Existing literature on this subject typically relies on simulating WDNs with pump scheduling methods to find an approximation of the optimal PV quantity, aiming to minimize the total cost of WDNs over the lifespan of PVs \cite{wdspv,MERCEDESGARCIA2022853}. However, these pump scheduling methods are often non-adaptive, meaning they cannot adjust pump schedules based on available PV forecasts and electricity prices to optimize WDN operating costs. Therefore, if a facility plans to use adaptive methods for operating its PV-powered WDN, depending on PV panel quantity recommendations from methods that simulate the network with non-adaptive pump scheduling may yield unreliable suggestions. This paper presents a procedure for the determination of an approximate value of the optimal quantity of PV panels for a specific WDN by optimizing the total cost of the network throughout the lifespan of PV panels through the use of an adaptive pump scheduling simulation.

To calculate the total cost of the WDN, it needs to be simulated using a pump scheduling method. The design of a pump scheduling method for WDNs is primarily defined by the considerable sizes of the networks and the presence of nonlinear network elements. Most of the available works in the literature can be classified based on their strategies for addressing the challenges posed by these nonlinearities and network sizes. Heuristic methods are used in some works to avoid running gradient-based optimization algorithms with complex models of the networks and nonlinear constraints. In \cite{SimBased}, the hydraulic simulation software EPANET is employed to generate and validate feasible solutions within a heuristic approach. Both \cite{Chase2009OptimalPS} and \cite{CastroGama2017PumpSF} use Genetic Algorithm (GA) with \cite{Chase2009OptimalPS} using pump operation times as decision variables and \cite{CastroGama2017PumpSF} using binary variables for pump opening and closing.

Some works have tried simplifying the network model often by relaxing the nonlinear pipe equations. In \cite{Baunsgaard2016MPCCO}, the approach involves approximating the pipe equations as linear equations and applying Model Predictive Control (MPC). The same linearization strategy is used in \cite{9372197} for a PV-powered WDN.  Similarly, an Economic Model Predictive Control (EMPC) method is developed with a set of linear inequalities derived from the relaxation of the nonlinear pipe equations \cite{Wang2018EconomicMP}. Another study, \cite{Fiedler2020EconomicNP},  adopts a different methodology by simplifying the network structure through node clustering. Each cluster is then represented by a single node, and a system model is developed using a Deep Neural Network (DNN) structure based on this simplified configuration. In \cite{Goryashko2014RobustEC}, the authors tackle the pump scheduling problem under uncertain water demand by introducing a Linear Programming framework and employing a robust methodology to handle the associated uncertainties. In \cite{Kallese2017PlugandPlayMP}, an EMPC is applied to a reduced model of the network. The EMPC formulation does not consider node pressure constraints. This omission is based on the assumption that the network has an elevated tank, and as long as the tank water levels are maintained above a certain threshold, the resulting pressure levels will remain within acceptable bounds. In this work, the same assumption is adopted. In recent works \cite{10383248,CCTAPaper}, EMPC-based pump scheduling methods are proposed for WDNs powered by grid-connected PV panels. The main difference between the method given in \cite{10383248} and the one in \cite{CCTAPaper}, which is also used in this paper, is the cost functions used in the controller calculation. The method in \cite{10383248} uses a deterministic cost function, while the one in \cite{CCTAPaper} uses a stochastic cost function to account for potential future scenarios in PV production.

The use of PV panels in any system requires the ability to predict their power production due to the volatility in power production. The PV power prediction problem is studied extensively in the literature with various prediction horizons, and types of data such as weather measurements, historical power production data, and Numerical Weather Predictions (NWP). NWP is employed alongside physical models of PVs to estimate power production using a closed-form equation, incorporating weather variables like radiance, temperature, and wind speed \cite{HULD20113359,8810897}. Some studies categorize days as sunny, cloudy, or rainy using NWP and adjust the predictions accordingly  \cite{THEOCHARIDES2020115023, Mellit2014ShorttermFO}. Some works solely rely on historical power data for their predictions. In \cite{Ordiano2017PhotovoltaicPF}, the use of NWP is avoided due to concerns about its availability in all locations. For short-term predictions, historical power data takes precedence as it is considered a more significant source of information, particularly for predictions within a few hours \cite{BACHER20091772,XIA2023110037}. In \cite{nwpcompare}, a comparative analysis was conducted between an NWP-based physical model and several NWP-independent methods for predicting the day-ahead power values. The findings revealed that the non-NWP-based methods performed better predicting the power values around noon compared to the NWP-based physical model.

Researchers have been looking into how to figure out the right number of PV panels to keep costs down in various systems \cite{anaf}, \cite{ZHANG2020115106}, \cite{oaos}, \cite{lsss}, \cite{dool}. In the context of WDNs powered by PVs, studies fall into two main types: those using grid-connected PV panels and those using off-grid solar systems. For off-grid systems, the main challenge is finding the amount of solar power needed to meet the network's energy demands \cite{PARDO2024119822}. In \cite{sysAmount}, a system with an Energy Storage System (ESS) is considered and the average yearly energy production of PVs is used to calculate the PV amount for WDNs. In \cite{doubleObj}, an approximation of the optimal PV size is determined using the firefly algorithm, with the constraint that daily PV energy output exceeds the daily energy required by water pumps to meet water demand, as the system did not have an ESS.  In \cite{BOUZIDI20131}, the PV amount is determined to minimize the operational costs while maintaining a reliability threshold,  measured as total delivered water divided by total water demand. For grid-connected systems, the challenge is a bit different. Now, the total energy from solar panels doesn't have to match the energy needed by WDNs because grid energy can help out. The goal is to find the right amount of PV panels to minimize costs over the life of the panels. In \cite{MERCEDESGARCIA2022853}, an approximation of the optimal PV amount is found for a grid-connected PV-powered irrigation system by simulating the system with an irrigation scheduling method. Another study \cite{wdspv} looked at designing solar-powered WDNs and optimized the number of panels to minimize total operating costs using a predetermined pump scheduling strategy. It's worth noting that in both \cite{wdspv} and  \cite{MERCEDESGARCIA2022853}, the pump schedules were not updated during simulations to minimize operating costs of the WDN, leading to a significant reduction in simulation time. However, if a facility aims to minimize costs by installing a minimal number of PVs and implementing adaptive pump scheduling methods, the recommendations from these approaches may be unreliable, as they do not incorporate adaptive pump scheduling methods.

\begin{figure*}[t]
\resizebox{1.4\columnwidth}{!}{\includegraphics {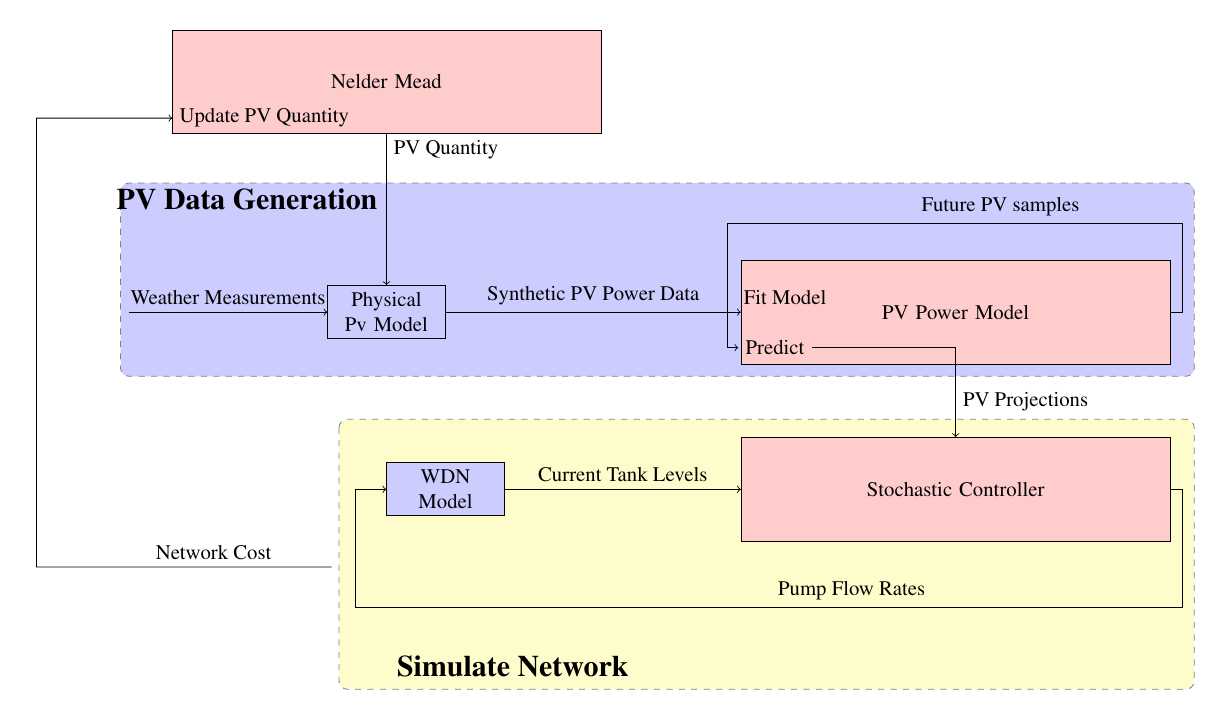}}
\centering
\caption{Overall method for optimizing the PV Quantity. The Nelder Mead method is used to update the PV quantity iteratively. For each PV quantity, a synthetic historical PV power data set is generated using a physical PV model and historical weather measurements. The PV power model is then fitted to the synthetic data, and PV samples are generated from this model. An approximate WDN model (see \eqref{eq:reducedModel}.) is simulated over the duration of PV samples treating the PV samples as the actual PV power generations powering the network. To simulate the network a stochastic controller-based pump scheduling method is used. Since future PV power values are unknown when pump schedules are calculated, the projections of future PV power production obtained from the PV model are used by the stochastic controller. Following the simulations, the total cost including the costs of power bought from the grid and PV installation, and maintenance costs throughout the lifespan of PVs is calculated. The PV quantity is updated according to the total cost using the Nelder Mead method.}
\label{fig:overall}
\end{figure*}

This paper presents two primary contributions. The first contribution is offering a systematic approach to determine an approximation of the optimal amount of PV panels to be installed for the desired pumping stations in a given WDN. {The objective is to find the PV panel quantity that optimizes the network's overall cost throughout the lifespan of the PV panels, encompassing installation, maintenance, and water pump usage costs.} This objective is chosen to enhance the appeal of installing PV panels in WDNs. The calculation of the total cost for a given amount of PV panels is done by generating samples of PV power production with a new probabilistic PV power production model and simulating the network using these samples in conjunction with a stochastic predictive controller based pump scheduling method introduced in \cite{CCTAPaper} because of its ability to consider the uncertainties associated with PV power production when establishing pump schedules. Iterative updates to the PV panel quantities are performed using a derivative-free optimization algorithm to find an approximation of the optimal number of PV panels, minimizing the total cost. The choice of a derivative-free method stems from the unavailability of gradient information for the total WDN costs. This is due to the fact that the total cost depends on calculating pump schedules through an optimization problem. Unlike prior research focused on optimizing the quantity of PV panels, the approach proposed in this paper involves using an adaptive pump scheduling method that tailors pump schedules based on the PV forecasts available and the electricity prices to minimize economic costs. Therefore,  this method can provide recommendations on the quantity of PV panels for facilities that want to use adaptive pump scheduling methods in the future. The proposed method is implemented on a linear model representing a medium-sized Danish town's network, specifically the Randers network. The flowchart of the optimization procedure is given in Figure \ref{fig:overall}. 

The second contribution is introducing a new probabilistic PV power production model. Building upon the method introduced in \cite{9968709} and refined in \cite{CCTAPaper}, the improvements include using a multiplicative correction term (see \eqref{eq:pvModel}.) instead of an additive one a given in \cite{CCTAPaper} to better model the intraday fluctuations in PV power production due to weather conditions. Additionally, the process of sampling future PV power has been improved compared to the one in \cite{CCTAPaper}, where the probability densities of the additive correction terms have not been updated with the arrival of the new PV values. In this paper, incoming PV values are used to update the densities of multiplicative correction terms. Finally, different stochastic models than those in \cite{9968709} are used in this paper to predict specific variables in the framework, which helps in the accurate modeling of PV data from locations where there are significant seasonal changes in solar power production.

The outline of the rest of the paper is as follows. PV power production model is given in Section \ref{sec:PvPow}. The model of the network is derived in Section \ref{sec:network}. The control method is explained in Section \ref{sec:control}. The procedure for optimizing the PV amount is explained in Section \ref{sec:Pvopt}. The experimental results are presented in Section \ref{sec:application}. The paper is concluded in Section \ref{sec:conc}.

\subsection{Notation}
The time derivative of a function $f(t)$ is shown with $\dot{f}(t)$. The sign $\odot$ is used to denote the elementwise multiplication of two matrices. The probability density function of a random variable $X$ is shown with $f_X(.)$. The conditional density of $X$ given the value $y$ of the random variable $Y$ is written as $f_{X \mid Y}(. \mid y)$. The Pontryagin difference is denoted with $\ominus$, where $A \ominus B=\{a \mid a+B \subseteq A \}$.

\section{PV Power Production Model} 
\label{sec:PvPow}
 In this section, the PV power production model is introduced, which is based on a method outlined in \cite{9968709}. The model serves two main purposes. First, it is used to generate future PV profile samples for a given number of PVs, enabling the simulation of the WDN with these samples as its power source. Second, the model is used to create realizations of future PV power production with a maximum one-day horizon. This helps in approximating the expected value in the cost function (see \eqref{eq:costMPC}.)  for the pump scheduling method. To fit the model parameters for the location where PVs are intended to be installed, the historical PV power production data is needed. However, assuming the availability of such historical data is not feasible, as the goal is to determine an approximation of the optimal PV installation without any PVs currently in place. To address this challenge, a synthetic historical PV power dataset is generated by utilizing historical weather measurements such as irradiance, temperature, and wind speed. These measurements are inputted into physical models that approximate PV power output as a function of weather variables. With the synthetic historical PV power data in hand, one can construct the PV power production model and fit its parameters using this synthetic historical data set. One might question the necessity of having a distinct model when physical models are already available. The reason is that those physical models won't work for the problem considered in this paper. A model capable of creating PV samples for a long time into the future, one year for the simulations in this paper, is required. However, the existing physical models require weather predictions for that duration, which is not available. Also, the model needs to be able to represent PV power probabilistically. This is necessary to generate realizations of future PV power with a one-day horizon, which are needed by the pump scheduling method.
\subsection{Synthetic Power Data}
To generate synthetic PV power data, the model presented in \cite{HULD20113359} is used, which is given by
\begin{equation}
\begin{split}
\label{eq:physPV}
P(G',T')=&G'(P_{STC}+k_1\ln(G')+k_2\ln(G')^2+k_3T' \\
&+k_4T'\ln(G')+k_5T'\ln(G')^2+k_6T'^2)
\end{split}
\end{equation}
where $k_i, i=1 \cdots 6$ are constants to be estimated for each PV panel type, $G', T'$ are the normalized irradiance and module temperature values and given by
\begin{subequations}
\label{eq:physNorm}
\begin{align}
&G'=G/G_{STC} \\
&T'=T_{mod}-T_{STC}
\end{align}
\end{subequations}
with $G_{STC}=1000 \: W/m^2, \:T_{STC}=25\,^\circ C$ being the irradiance and temperature values under standard conditions. $P_{STC}$ is the power production under normal conditions i.e. $G=G_{STC},T_{mod}=T_{STC}$.  The temperature of the PV module $T_{mod}$ can be approximated using the Faiman model presented in \cite{faiman} as
\begin{align}
    \label{eq:ModTemp}
    T_{mod}=T_{amb}+\frac{G}{\mu_0+\mu_1V_{wind}}
\end{align}
where $T_{amb}$ is the ambient temperature $V_{wind}$ is the wind speed and $\mu_0,\mu_1$ are constant parameters. To estimate the parameters $k_1-k_6,\mu_0,\mu_1$, a small number of PVs should be installed and the irradiance, temperature, and power measurements must be taken for a while. Then, synthetic historical PV power data can be generated using historical weather measurements.
\subsection{PV Power Model}
 The challenge in modeling PV power generation is that it has both seasonal and daily patterns, along with intraday fluctuations.  The sun's daily trajectory relative to PV panels changes throughout the year. Consequently, sunrise and sunset times, as well as the PV production at each hour of the day compared to other hours, undergo seasonal changes. The angle of sunlight also changes throughout the year, impacting the average daily PV production. Weather conditions contribute to the complexity by causing intraday fluctuations in PV production. In this section, a PV model capturing these characteristics is presented. The PV power production is modeled as
\begin{align}
    \label{eq:pvModel}
    X_{\eta}=p_{\eta}Y_{\eta}\odot \delta_{\eta}
\end{align}
where $ X_{\eta}\in\mathbb{R}^{N_{pv}}$ is the power output of the PVs at day $\eta$, $N_{pv} \in\mathbb{R}_{>0}$ is the number of data points in a day, $p_{\eta}\in\mathbb{R}$ is the optimal multiplier (see \eqref{eq:optMultDef}.), $Y_{\eta}\in\mathbb{R}^{N_{pv}}$ is the daily power profile and $\delta_{\eta}\in\mathbb{R}^{N_{pv}}$ is the multiplicative correction term (see \eqref{eq:multCorrDef}.). Note that multiplicative correction terms are used in this paper instead of the additive ones given in \cite{CCTAPaper}. The purpose of the model is to enable day-ahead power forecasting and generate samples of PV profiles.

\begin{figure}[bt]
\centering
{\includegraphics[width=0.4\textwidth]{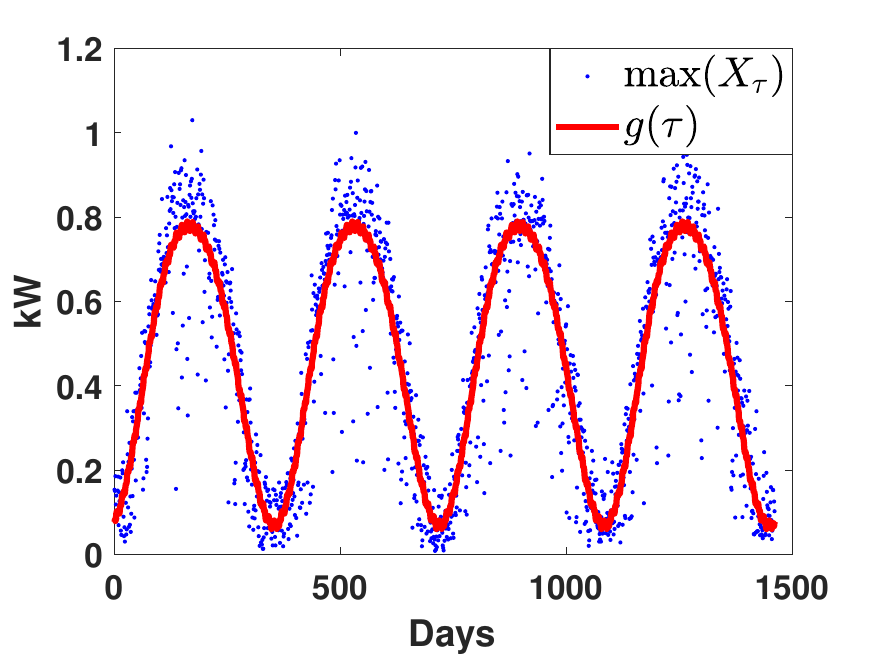}}
\caption{An illustration of the $g(\tau)$ fitted to the time series of the daily maximum power production $\max(X_{\tau})$ of the synthetic PV data generated with $P_{STC}=1$ kW.}
\label{fig:maxValues}
\end{figure}

The daily power profile $Y_{\eta}$ represents the change of the sun's position in a day and provides information about the relative amount of irradiation received at each time during the day. Since the sun's intraday position changes are very similar between consecutive days, $Y_{\eta}$ is calculated using the Exponentially Weighted Moving Average (EWMA) method as 
\begin{subequations}
\label{eq:EWMA}
\begin{align}
%\begin{split}
 X'_{\eta-1} &= \frac{1}{g(\eta-1)}  X_{\eta-1},
\\
\label{eq:yUpdate}
Y_\eta &= \alpha X'_{\eta-1} + (1-\alpha)Y_{\eta-1}
%\end{split}
\end{align}
\end{subequations}
where $g(\eta-1)$ is a smoothed form of the daily maximum power production $\max(X_{\eta-1})$. This smoothing is achieved by fitting a one-year periodic function to the daily maximum production data $\max(X_{\tau}), ~ \tau=1,\dots,N_{s}$ of synthetic PV data using least squares, where $N_{s}$ is the number of days in the synthetic dataset. An illustration of the function $g(\tau)$ along with the corresponding daily maximum power production is given in Figure \ref{fig:maxValues}. The power profile $Y_{\eta}$ is then obtained by using EWMA \eqref{eq:yUpdate} with the normalized daily data $X'_{\eta-1}\in\mathbb{R}^{N_{pv}}$. The EWMA parameter $\alpha \in [0,1]$ governs the influence of recent days relative to earlier ones in shaping $Y{\eta}$. The rationale behind using normalized daily data $X'$ instead of $X$ in the computation of the next day's daily profile  $Y_{\eta}$ is to balance the influence of days characterized by high PV production and those with low PV production on the values of $Y_{\eta}$. 

The optimal multiplier $p_{\eta}$ is defined as
\begin{subequations}
\label{eq:optMultDef}
\begin{align}
p_{\eta}&=\argmin_{p} \quad  \sum_{i=1}^{{N_{pv}}}{(pY_{\eta }^i-X_{\eta }^i)^2}\label{eq:optim1} \\
 &= \frac{\sum_{i=1}^{{N_{pv}}}{Y_{\eta}^iX_{\eta}^i} }{\sum_{i=1}^{{N_{pv}}}{{Y_{\eta }^i}^2}}.\label{eq:optim2}
\end{align}
\end{subequations}
where $Y_{\eta }^i$ (resp. $X_{\eta }^i$) are the $i^{th}$ coordinate of $Y_\eta$ (resp. $X_\eta$). As it is the multiplier of $Y_{\eta }$ such that $p_{\eta}Y_{\eta }$ is closest to the real production data $X_{\eta }$, it can be thought as representing the average daily weather condition which gets bigger values on clear days, and smaller values on cloudy days. Calculating the optimal multiplier $p_\eta$ relies on having all the power values for the current day $X_\eta$ accessible through equation \eqref{eq:optim2}. Consequently, to estimate the power values $X_\eta$ before the day $\eta$, a probabilistic model for the optimal multiplier $p_\eta$ is needed. Optimal multipliers are modeled as
\begin{align}
\label{eq:opmultmodel}
p_{\eta}&=(\gamma({\eta})+\epsilon_{p_{\eta}})^2
\end{align}
where $\gamma({\eta})$ is the term representing the seasonal changes of optimal multipliers, and $\epsilon_{p_{\eta}}$ is the error term representing the daily deviations from the seasonal changes. Please note that the original model introduced in \cite{9968709} didn't include a seasonal term for the optimal multipliers, so it couldn't account for seasonal changes in PV production. The seasonal part $\gamma({\eta})$ is obtained by fitting a periodic function with a period of one year to the time series of the square root of optimal multipliers $\sqrt{p_{\tau}}, ~ \tau=1,\dots, N_s$ of the synthetic dataset using least squares. The error terms $\epsilon_{p_{\tau}} ~ \tau=1,\dots,N_s$ are then obtained as $\epsilon_{p_{\tau}}=\sqrt{p_{\tau}}-\gamma({\tau})$. Since the optimal multiplier is a nonnegative variable, it is modeled as the square of a random variable so that when $\epsilon_{p_{\eta}}$ is sampled from its respective distribution the resulting optimal variable ${p_{\eta}}$ is always nonnegative. The error values $\epsilon_{p_{\eta}}$ are modelled using an Autoregressive Moving Average model, ARMA(1,1), as 
\begin{align}
\label{eq:optMultErrMod}
\epsilon_{p_{\eta}}&=\mu_p + \phi_p \epsilon_{p_{\eta-1}} + \theta_p \zeta_{p_{\eta-1}} + \zeta _{p_{\eta}}
\end{align}
where $\zeta_{p_{\eta}}$ and $\zeta_{p_{\eta-1}}$ are the forecast errors at step $\eta, \eta-1$ respectively and $\mu_p, \phi_p, \theta_p$ are the constants whose values are found by fitting the time series $\epsilon_{p_{\tau}}, ~ \tau=1,\dots,N_s$ to the model. The error terms $\zeta_{p_{\eta}}$ are assumed to be coming from independent zero-mean normal distributions.

Multiplicative correction terms $\delta_{\eta}$ are defined as
\begin{align}
\label{eq:multCorrDef}
    \delta_{\eta}^i=\frac{X_\eta^i}{p_\eta Y_\eta^i}.
\end{align}
They capture the intraday changes in weather conditions. If the weather remains stable throughout a day either clear or cloudy, correction terms $\delta_{\eta}$ stay around the value $1$. On the other hand, if there is an inconsistency in the weather conditions for example a clear sky at time instance $i$ on a cloudy day $\eta$, the correction term $\delta_{\eta}^i>1$ attains a bigger value. The choice of multiplicative correction terms, as opposed to the additive ones used in \cite{CCTAPaper}, aims to create a correction term time series whose statistical properties such as mean and variance do not vary a lot in time so that it can be better modeled with a stationary process. When there's a change in weather conditions during high PV production, the impact on PV power production is more significant compared to periods of low production. For instance, a sudden cloud cover would cause a greater reduction in PV power production at noon than near sunset. In an additive correction term model, this would lead to larger correction term values at noon and smaller ones away from noon. However, in the multiplicative correction term model, the values of correction terms at noon and during other times are closer to each other. Similar to optimal multipliers, a probabilistic model for correction terms $\delta_{\eta}^i$ is necessary to estimate PV power $X_\eta$, since these terms are determined by the power values of the current day according to equation \eqref{eq:multCorrDef}. The correction terms at night are insignificant since both the power production $X_\eta$ and the daily power profile $Y_\eta$ values are zero. Therefore, only the correction terms during the day are modeled as
\begin{align}
\label{eq:multErr}
    \ln\delta_{\eta}^i=\mu_{\delta}+\phi_{\delta} \ln\delta_{\eta}^{i-1}+\zeta _{\delta_{\eta}}^i
\end{align}
where $\mu_{\delta},\phi_{\delta}$ are constant parameters and $\zeta _{\delta_{\eta}}^i$ are assumed to be coming from independent zero-mean normal distributions. Given that the correction terms $\delta_{\eta}^i$, as defined in  \eqref{eq:multCorrDef}, are inherently positive, using the logarithmic function $\ln$ to represent these correction terms in the model \eqref{eq:multErr} ensures that the predictions for the correction terms will also remain positive. This model is developed solely based on the historical time series of the correction terms. However, when the previous day $\eta-1$ ends and the daily power profile $Y_\eta$ is obtained using \eqref{eq:EWMA}, it tells more about $\delta_{\eta}$ which could be used to update the distributions of the $\zeta _{\delta_{\eta}}^i$ terms. Specifically, multiplicative correction terms should satisfy a certain equation which could be derived as
\begin{align}
    \nonumber
    p_{\eta}&=\frac{\sum_{i=1}^{{N_{pv}}}{Y_{\eta}^iX_{\eta}^i} }{\sum_{i=1}^{{N_{pv}}}{{Y_{\eta }^i}^2}}
    \\
    \nonumber
    &=\frac{\sum_{i=1}^{{N_{pv}}}{Y_{\eta}^ip_{\eta}Y_{\eta}^i\delta_\eta^i} }{\sum_{i=1}^{{N_{pv}}}{{Y_{\eta }^i}^2}}
    \\
    \sum_{i=1}^{{N_{pv}}}{{Y_{\eta }^i}^2}\delta_\eta^i &=\sum_{i=1}^{{N_{pv}}}{{Y_{\eta }^i}^2}.
    \label{eq:corrUpdate}
\end{align}
It can be seen that the true values of the correction terms satisfy this equation by substituting $\delta_\eta^i$ with its definition \eqref{eq:multCorrDef} into \eqref{eq:corrUpdate}. This substitution yields the definition of the optimal multiplier \eqref{eq:optMultDef}. That means the variable definitions in the PV model are consistent with \eqref{eq:corrUpdate}. After the preceding day $\eta-1$ and the acquisition of the daily power profile $Y_\eta$, the correction terms $\delta_\eta^i$ remain unknown. In the absence of \eqref{eq:corrUpdate}, the correction term model \eqref{eq:multErr} would be used for estimating these terms. However, incorporating the information from \eqref{eq:corrUpdate} with the correction term model \eqref{eq:multErr} following the arrival of $Y_\eta$ allows for a more accurate estimation of the correction terms. With this information, variables $\zeta _{\delta_{\eta}}^i$ are no longer independent from each other. Therefore, when the value of the daily power profile $Y_\eta$ becomes available, and \eqref{eq:corrUpdate} can be used, the initial distributions of $\zeta _{\delta_{\eta}}^i$, which are independent zero-mean normal distribution, cannot be used to sample or estimate them. Instead, the probability densities of the $\zeta _{\delta_{\eta}}^i$ terms are updated at the beginning of the day $\eta$ using \eqref{eq:corrUpdate}  as
\begin{align}
    &f_{\zeta _{\delta_{\eta}}^{sr},\zeta _{\delta_{\eta}}^{sr+1}\cdots \zeta _{\delta_{\eta}}^{sd} \mid \sum_{i=1}^{{N_{pv}}}{{Y_{\eta }^i}^2}\delta_\eta^i}(Z _{\delta_{\eta}}^{sr},Z _{\delta_{\eta}}^{sr+1}\cdots Z _{\delta_{\eta}}^{sd} \mid   \sum_{i=1}^{{N_{pv}}}{{Y_{\eta }^i}^2}) \propto \nonumber \\ 
    \label{eq:updErr}
    & \begin{cases}
    f_{\zeta _{\delta_{\eta}}^{sr}}(Z _{\delta_{\eta}}^{sr})\cdots f_{\zeta _{\delta_{\eta}}^{sd}}(Z _{\delta_{\eta}}^{sd}),& \sum_{i=1}^{{N_{pv}}}{{Y_{\eta }^i}^2}\delta_\eta^i(Z) =\sum_{i=1}^{{N_{pv}}}{{Y_{\eta }^i}^2}\\
    0,              & otherwise
\end{cases} 
\end{align}
where $sr,sd$ are the time indices for sunrise and sundown respectively and $\delta_\eta^i(Z), ~ i=sr,\cdots sd$ are the correction terms obtained by substituting the sampled values $Z _{\delta_{\eta}}^{i}, ~ i=sr,\cdots sd$ in place of the corresponding $\zeta _{\delta_{\eta}}^i$ in the correction error term model \eqref{eq:multErr}. This probability update essentially results from the combination of the general information concerning the correction terms\eqref{eq:multErr} and the new information \eqref{eq:corrUpdate} obtained at the beginning of the day $\eta$. Before obtaining the daily power profile $Y_\eta$, the occurrence of any sequence of $Z _{\delta_{\eta}}^{i}, ~ i=sr,\cdots sd$ had a positive probability, since $\zeta _{\delta_{\eta}}^i$ terms are assumed to be coming from independent zero-mean normal distributions. Once $Y_\eta$ values become available, only those specific sequences of $Z_ {\delta_{\eta}}^{i}$ that yield $\delta_\eta^i(Z)$ values satisfying the condition \eqref{eq:corrUpdate} maintain a positive probability, while the remaining sequences get a zero probability. Note that in the model used in \cite{CCTAPaper}, the probability densities of the relevant variables are not updated using information obtained with the arrival of $Y_\eta$.

 In simple terms, the PV model has three parts representing different aspects of solar power production. The optimal multiplier $p_{\eta}$ indicates the level of sunlight on a given day. The change of position of the sun relative to PV panels is described by $Y_{\eta}$, and any deviations from the daily power production due to weather conditions are noted by $\delta_{\eta}$. Each of these elements has its own probability representation in the time series, allowing estimation of their values or generation of realizations of them when their exact values are unknown.
\subsection{PV power data sampling}
\label{sec:PvSample}
In this section, the process of sampling PV power from the PV power model is described. It's worth noting that the sampling method described in this section is different from the one presented in \cite{CCTAPaper} because of the introduction of multiplicative correction terms $\delta_\eta$. Another reason for this difference is that the probability densities of the relevant variables are not updated using \eqref{eq:corrUpdate} in \cite{CCTAPaper}.  Generating future PV power samples is needed both for the optimization of the PV amount as detailed in Section \ref{sec:Pvopt} and for the stochastic pump scheduling method presented in Section \ref{sec:control}.  In the context of PV amount optimization, it is necessary to generate future samples of PV power production for a specified number of PVs. These samples are used to simulate the WDN as if they represented the actual one-year PV power values fueling the entire WDN.  On the other hand, for the stochastic pump scheduling method, realizations of PV power production until the day's end, at each instance of pump schedule calculation, are used to estimate the expected value in the cost function (see\eqref{eq:costMPC}.). It's important to note that the samples created for the pump scheduling method are distinct from the one-year samples. This is because, in real-life situations, when we calculate pump schedules, we don't know the future values of these samples, and the same assumption is made in our simulations. The one-year samples are regarded as actual power production values, whereas the samples for the pump scheduling method are projections of these actual values for a one-day horizon before we observe them. The problem of sampling can be stated as follows. Let $i_c$ denote the current time index in day $\eta$. Then, the task is to sample realizations of the unseen power values of the current day $X_\eta^i, ~ i=i_c+1,i_c+1\cdots N_{pv}$. Once a sampling method is identified, obtaining one-year samples involves sequentially generating daily samples for the entire year. For the stochastic pump scheduling method, the sampling method could be used at each instance $i_c$ of pump schedule calculation to generate realizations of PV power generation until the end of the day. The sampling task exhibits a distinction between the period before sunrise and after sunrise due to the availability of power measurements. After sunrise, the power measurements $X_\eta^i, ~ i=0,1\cdots i_c-1$ become available at the time $i_c$ which can be used to update the probability distribution of the variables.

The values of $X^i_\eta$ are generated by sampling the optimal multiplier $p_\eta$ and the multiplicative correction terms $\delta_{\eta}^i$ since $Y_{\eta}$ is already calculated at the start of the day $\eta$. Before sunrise, $p_\eta$ values are generated by sampling $\epsilon_{p_{\eta}}$ from its respective probability distribution and putting the sampled value to
 the optimal multiplier model \eqref{eq:opmultmodel}. The samples of multiplicative correction values $\delta_{\eta}^i$ are generated by sampling $\zeta _{\delta_{\eta}}^{i}$ according to updated distribution \eqref{eq:updErr} and putting those samples into the model of correction terms \eqref{eq:multErr}. To sample $\zeta _{\delta_{\eta}}^{i}$ values from the updated distributions \eqref{eq:updErr}, a rejection sampling method is employed. The $\zeta_{\delta_{\eta}}^{i}$ values are generated independently from their normal distributions, and the samples are accepted if the resulting $\delta_\eta^i(Z)$ values, derived by substituting the sampled $\zeta_{\delta_{\eta}}^{i}$ values into the correction term model \eqref{eq:multErr}, satisfy the following inequality
\begin{align}
    \label{eq:sampleRule}
    -\epsilon<\sum_{i=1}^{{N_{pv}}}{{Y_{\eta }^i}^2}\delta_\eta^i -\sum_{i=1}^{{N_{pv}}}{{Y_{\eta }^i}^2}<\epsilon
\end{align}
for some small $\epsilon \in \mathbb{R}_{>0}$. The samples obtained using this method closely approximate the updated distributions \eqref{eq:updErr}.

After sunrise, non-zero power production data $X^i_\eta$ starts coming in, giving more information about the variables $\delta^{i}_\eta$ and $p_\eta$. Let the index corresponding to sunrise be denoted as $sr$. At time $i_c$ the $\delta^{i}_\eta, ~ i=sr+1,sr+2 \cdots i_c$ values become deterministic functions of $p_\eta$ given by the equations $\delta^{i}_\eta= X^i_\eta/(p_\eta Y^i_\eta)$ for $i=sr+1,sr+2 \cdots i_c$ as the values of $X^i_\eta, ~ i=sr+1,sr+2 \cdots i_c$ are available at time $i_c$. Then, 
the error values $\zeta _{\delta_{\eta}}^{i}= k_{i}(p_\eta)$ for $i=sr+1,sr+2 \cdots i_c$ also become deterministic functions of $p_\eta$, where $k_{i}(p_\eta)$ can be obtained by replacing the correction terms $\delta^{i}_\eta$ with $X^i_\eta/(p_\eta Y^i_\eta)$ in equation \eqref{eq:multErr}. Given this information, the probability density of $p_\eta$ is updated using
\begin{subequations}
\begin{align}
\label{eq:sampleCond}
&f_{p_\eta \mid \zeta _{\delta_{\eta}}^{sr+1},\cdots,\zeta _{\delta_{\eta}}^{i_c},\sum_{i=1}^{{N_{pv}}}{{Y_{\eta }^i}^2}\delta_\eta^i}(x \mid k_{sr+1}(p_\eta), \\
&\cdots,k_{i_c}(p_\eta),\sum_{i=1}^{{N_{pv}}}{{Y_{\eta }^i}^2} ) \nonumber \\ 
\label{eq:sampleBayes}
&\propto f_{p_\eta}(x)f_{\zeta _{\delta_{\eta}}^{sr+1},\cdots,\zeta _{\delta_{\eta}}^{i_c},\sum_{i=1}^{{N_{pv}}}{{Y_{\eta }^i}^2}\delta_\eta^i \mid p_\eta }(k_{sr+1}(p_\eta),  \\
&\cdots,k_{i_c}(p_\eta),\sum_{i=1}^{{N_{pv}}}{{Y_{\eta }^i}^2} \mid x ) \nonumber \\ 
\label{eq:sampleSubs}
&=f_{p_\eta}(x)f_{\zeta _{\delta_{\eta}}^{sr+1},\cdots \zeta _{\delta_{\eta}}^{i_c}, \sum_{i=1}^{{N_{pv}}}{{Y_{\eta }^i}^2}\delta_\eta^i}(k_{sr+1}(x),\cdots k_{i_c}(x),  \\
& \sum_{i=1}^{{N_{pv}}}{{Y_{\eta }^i}^2}) \nonumber \\
\label{eq:sampleCond2}
&= f_{p_\eta}(x)f_{\zeta _{\delta_{\eta}}^{sr+1},\cdots \zeta _{\delta_{\eta}}^{i_c}}(k_{sr+1}(x),\cdots k_{i_c}(x)) \\
&f_{\sum_{i=1}^{{N_{pv}}}{{Y_{\eta }^i}^2}\delta_\eta^i \mid \zeta _{\delta_{\eta}}^{sr+1},\cdots \zeta _{\delta_{\eta}}^{i_c}}(\sum_{i=1}^{{N_{pv}}}{{Y_{\eta }^i}^2} \mid k_{sr+1}(x),\cdots k_{i_c}(x)) \nonumber \\
\label{eq:sampleFinal}
&=f_{p_\eta}(x)f_{\zeta _{\delta_{\eta}}^{sr+1}}(k_{sr+1}(x))\cdots f_{\zeta _{\delta_{\eta}}^{i_c}}(k_{i_c}(x)) \\
&f_{\sum_{i=i_c+1}^{{N_{pv}}}{{Y_{\eta }^i}^2}\delta_\eta^i }(\sum_{i=1}^{{N_{pv}}}{{Y_{\eta }^i}^2}-\sum_{i=1}^{i_c}{{Y_{\eta }^i}^2}X_\eta^{i}/(xY_{\eta }^i)). \nonumber
\end{align}
\end{subequations}
The relation \eqref{eq:sampleBayes} is obtained by applying Bayes' rule $f_{X \mid Y=y}(x) \propto f_X(x)f_{Y \mid X=x}(x)$ to \eqref{eq:sampleCond}. Then, the conditional probability in \eqref{eq:sampleBayes} is transformed to unconditional probability \eqref{eq:sampleSubs} by setting $p_\eta=x$. Then, the conditional density rule $f_{X,Y}=f_Y(y)f_{X \mid Y=y}(x)$ is applied to get \eqref{eq:sampleCond2}.
 Finally,  the independency of $\zeta _{\delta_{\eta}}^i$ terms are used to separate joint probabilities $f_{\zeta _{\delta_{\eta}}^{sr+1},\cdots \zeta _{\delta_{\eta}}^{i_c}}(.)$ into marginal distributions $f_{\zeta _{\delta_{\eta}}^{i}}(.)$ in \eqref{eq:sampleFinal}. The conditional probability $f_{\sum_{i=1}^{{N_{pv}}}{{Y_{\eta }^i}^2}\delta_\eta^i \mid \zeta _{\delta_{\eta}}^{sr+1},\cdots \zeta _{\delta_{\eta}}^{i_c}}$ in \eqref{eq:sampleCond2} is transformed into the unconditional version in \eqref{eq:sampleFinal} by setting $\delta^{i}_\eta= X^i_\eta/(x Y^i_\eta)$ for $i=sr+1,sr+2 \cdots i_c$ as the conditions $\zeta _{\delta_{\eta}}^{i}= k_{i}(p_\eta)$ for $i=sr+1,sr+2 \cdots i_c$ are equivalent to $\delta^{i}_\eta= X^i_\eta/(x Y^i_\eta)$. 
 
 The summation $\sum_{i=i_c+1}^{{N_{pv}}}{Y_{\eta }^i}^2\delta_\eta^i$ in \eqref{eq:sampleFinal} is a linear combination of random variables ${Y_{\eta }^i}^2\delta_\eta^i$ that follow a lognormal distribution as deduced from the model of the correction terms \eqref{eq:multErr}. To approximate the probability density of this summation, a method presented in \cite{SumlogNormal} is employed. This method allows for the approximation of the probability density of the sum of two correlated lognormal random variables with a lognormal distribution. Then, the optimal multiplier can be sampled from the updated density \eqref{eq:sampleFinal}. The next task is to sample the corrections terms $\delta^{i}_\eta, ~ i=i_c+1,i_c+2 \cdots sd$ to get the sampled power values $X_\eta^i ~ i=i_c+1,i_c+2 \cdots sd$, where $sd$ is the index for sundown.  Given the optimal multiplier sample $p_\eta=x$, the probability densities of the $\zeta _{\delta_{\eta}}^i$ terms are updated as 
\begin{align}
    &f_{\zeta _{\delta_{\eta}}^{i_c+1},\zeta _{\delta_{\eta}}^{i_c+2}\cdots \zeta _{\delta_{\eta}}^{sd} \mid p_{\eta}}(Z _{\delta_{\eta}}^{i_c+1},Z _{\delta_{\eta}}^{i_c+2}\cdots Z _{\delta_{\eta}}^{sd} \mid  x) \propto \nonumber \\ 
    \label{eq:updErr2}
    & \begin{cases}
   f_{\zeta _{\delta_{\eta}}^{i_c+1}}(Z _{\delta_{\eta}}^{i_c+1})\cdots f_{\zeta _{\delta_{\eta}}^{sd}}(Z _{\delta_{\eta}}^{sd}),& \sum_{i=1}^{{N_{pv}}}{{Y_{\eta }^i}^2}\delta_\eta^i =\sum_{i=1}^{{N_{pv}}}{{Y_{\eta }^i}^2}\\
    0,              & otherwise
    \end{cases}
\end{align}
where $\delta_\eta^i= X^i_\eta/(x Y^i_\eta)$ for $i=sr,\cdots i_c$.  The $\zeta_{\delta_{\eta}}^i$ values are sampled using the rejection sampling method from the updated density \eqref{eq:updErr2}, similar to the way they were sampled before sunrise.
\section{Water Distribution Network Model}
\label{sec:network}
In this section, a control-oriented model of WDNs is presented. A water distribution network is composed of various components such as pipes, demand nodes, pumps, tanks, and more. In this paper, WDNs with elevated tanks either directly connected or standing alone are considered. It is assumed that the network has elevated tanks, and as long as the tank water levels are maintained above a certain threshold, the resulting pressure levels in the network will remain within acceptable bounds with the help of valves. Additionally, pumps in the network are considered to be variable-speed pumps and each pumping station is assumed to have pump inverters letting the facility control the flow rate of the pumps. The hydraulic head is a measure of fluid pressure, representing the height of a static fluid column at a specific point. The loss of hydraulic head in a pipe can be estimated using the Hazen-Williams Equation as
\begin{equation}
\Delta h=h_1-h_2=Kq\lvert q \rvert ^{0.852} 
\label{eq:Haz-Will}
\end{equation}
where $K$ is the pipe resistance that depends on the physical features of a pipe such as the diameter and length of the pipe, $q$ is the flow rate, and $h_1$ and $h_2$ are the heads at the two ends of the pipe. At each node $j$,  the inflow of water is equivalent to the outflow of water, which can be expressed as  
\begin{equation}
 \sum_{i \in \mathpzc{N}_j} q_{ij} = d_j
 \label{eq:NodeMass}
\end{equation}
where $q_{ij}$ is the flow entering the node $j$ from node $i$ and $d_j$ is the demand at node $j$, which is the water requested by the user at node $j$. The symbol $\mathpzc{N}_j$ denotes the set of neighbor nodes of node $j$. Note that  $q_{ij}$ is positive if the flow is from node $i$ to the neighbor node $j$ and negative vice versa.

Valves are used to control the flow or the pressure at specific points in a WDN. There are various types of valves such as Pressure Reducing Valves (PRVs), and Flow Control Valves (FCVs). PRVs limit the pressure at a point in the WDN. FCVs limit the flow below a certain level. In this work, it is assumed that the valves in the network are controlled according to pre-defined strategies to manage the pressure and flow at specific locations.

The water tanks are the only dynamic components within a WDN, and their behavior can be described as
\begin{equation}
 A_j\dot{{h}}_j = \sum_{i \in \mathpzc{N}_j} q_{ij}
 \label{eq:tankLevel}
\end{equation}
where $A_j$ represents the cross-sectional area of the tank identified by index $j$, $h_j$ is the tank level, $q_{ij}$ is the flow entering the tank, and $\mathpzc{N}_j$ denotes the set of neighbor nodes of the tank $j$. Tank levels change according to the flow passing through the pipes connected to the tanks. These flow rates $q_{ij}$ are affected by the valve control methods and head loss equations \eqref{eq:Haz-Will} of the pipes in the network, and mass balance equations \eqref{eq:NodeMass} of the demand nodes in the whole network. The flow rates of the pipes connected to the tanks $q_{ij}$ can be written as functions of demand at each node, tank levels, and the amount of water coming from the pumps as
\begin{equation}
 q_{ij}=f_{ij}(h,q,d)
 \label{eq:generalflow}
\end{equation}
where $h\in \mathbb{R}^{n},q \in \mathbb{R}^{m}$ are the vectors containing the tank levels and pump flow rates respectively, and $d=[d_1, d_2 ...]^T$ is the vector containing the demands of all the nodes. Combining \eqref{eq:generalflow} with the tank level equations \eqref{eq:tankLevel} gives
\begin{equation}
 A_j\dot{{h}}_j = \sum_{i \in \mathpzc{N}_j} f_{ij}(h,q,d).
 \label{eq:tankgeneral}
\end{equation}
Note that the functions $f_{ij}$ are nonlinear because of the nonlinearity in the head loss equations \eqref{eq:Haz-Will} of the pipes and the potential nonlinearity introduced by the valves. Since not all facilities keep track of water demands at each node, it is assumed that such detailed data is unavailable. Instead, it is assumed that the total demand of the area served by the pumps can be estimated based on historical data. Consequently, the tank dynamics can only be approximated because of the lack of knowledge of individual demands. To approximate the tank dynamics described by equation \eqref{eq:tankgeneral}, a linear model is used and written as
\begin{equation}
 \dot{{h}}(t)= Ah(t) + B_1u(t) + B_2 d_{a}(t)
 \label{eq:reducedModel}
\end{equation}
where $h(t) \in \mathbb{R}^{n}$ includes tank levels measured in meters, $A \in \mathbb{R}^{n\times n}$, $B_1 \in \mathbb{R}^{n\times m}$, $B_2 \in \mathbb{R}^{n\times 1}$ are constant system matrices and $d_{a}(t)$ is the aggregated demand of the area supplied by the pumping stations at time $t$ measured in $m^3/s$, $u(t) \in \mathbb{R}^{m}$ is the input containing the flow rate of the water exiting the pumping stations  measured in $m^3/s$.  A linear model is chosen for control-oriented modeling because it reasonably approximates the tank level evolution given by the EPANET model and it is computationally efficient when used in optimization-based control methods (see \eqref{eq:MPCForm}). A more complex model can be built and that model would be more accurate than the linear model \eqref{eq:reducedModel}. However, with a complex model, the computation of the control input would last longer if the model were to be used with an optimization-based method. Therefore, the use of such a complex model to compute the control input in an online fashion may not have been efficient.  Note that the fast pipe dynamics in the network are not accounted for in our model. That is because tank dynamics are so slow that the effect of the fast pipe dynamics on the tank levels is negligible. The matrices $A, B_1, B_2$ are obtained by applying least squares to fit linear functions of tank levels, pump flows, and the aggregated demand to the flow rates of the pipes $q_{ij}$ connected to the pumps, using the data generated by the EPANET model of the WDN.
\section{Stochastic Economic Model Predictive Controller for
Pump Scheduling}
\label{sec:control}
This section presents a predictive control based pump scheduling method for WDNs that are powered by grid-connected PV panels. The approach, initially proposed in \cite{CCTAPaper}, focuses on minimizing the economic costs involved in WDN operation while satisfying network constraints. To implement this method, a central computer and a communication infrastructure linking tanks and pumps to this central unit are assumed prerequisites. Consequently, the central computer has access to tank levels and the ability to solve the control problem. Upon resolving the problem, it communicates the calculated flow rates to the pumping stations, where local controllers subsequently modify the pump flow rates accordingly. The problem at time $t$ is formulated as 
\begin{subequations}
\label{eq:MPCForm}
\begin{align}
 &\min_{u_0^t,u_1^t \cdots u_{N(t)-1}^t}  \sum_{j=0}^{N(t)-1} \E[J(h_j^t,u_j^t,t, P_{pv},c)] \label{eq:costMPC}
\\
 &h_j^t=A_dh_{j-1}^t+B_{d1}u_{j-1}^t+B_{d2} d^t_{a}(j-1)
 \label{eq:discState}
 \\
 &h_0^t=h(t)
\\
 &u_j^t  \in \mathcal{U} \subseteq \mathbb{R}^{m}&
 \\
 &h_j^t  \in \mathcal{H} \subseteq \mathbb{R}^{n}& \label{eq:MPCstateCons}
 \\
&h_{N(t)}^t \in \mathcal{H}_{tf} \subseteq \mathbb{R}^{n}
\end{align} 
\end{subequations}
where $\E[J(h_j^t,u_j^t,t,P_{pv},c)]$ is the expected value of economic cost function $J(h_j^t,u_j^t,t, P_{pv},c)$, $h^t=[h_1^t\cdots h_{N(t)}^t ] \in \mathbb{R}^{n\times N(t)}$ is the predicted future states, $u_j^t \in \mathbb{R}^{m}$ is the input vector, $d^t_{a}$ is the estimated aggregated demand of the area supplied by the pumping stations, $ P_{pv}$ is the PV power, $c$ is the electricity prices in $\text{\euro}/kW$, $N(t)$ is the prediction horizon, $\mathcal{U} \subseteq \mathbb{R}^{m}$ and $\mathcal{H} \subseteq \mathbb{R}^{n}$ denotes the input and state constraints respectively and $\mathcal{H}_{tf} \subseteq \mathbb{R}^{n}$ is the terminal state set. Equation \eqref{eq:discState} represents the discretized version of the continuous system \eqref{eq:reducedModel}. At each time step, which is separated by a time interval of $\Delta_t$, the optimization problem \eqref{eq:MPCForm} is solved and the first term $u_0^{t}$ of the optimal input sequence $\mathbf{u}^t=[u_0^{t} \cdots u_{N(t)-1}^t] \in \mathbb{R}^{m\times N(t)}$ is applied to the system. 

The minimum and maximum flow rate capacity of the pumps determines the input constraints for the system. These conditions are expressed as 
\begin{multline}
 \mathcal{U}=
 \{[u_1\cdots u_m]^T \in \mathbb{R}^{m} \mid \forall i : 0\leq u_i \leq \overline u_i\}
 \label{eq:inputConstr}
% \mathcal{U}=
% \{[u_1\cdots u_m]^T \in \mathbb{R}^{m} \mid 0\leq u_1 \leq \overline u_1, \cdots, 0\leq u_m \leq \overline u_m \}
% \label{eq:inputConstr}
\end{multline}
where $\overline u_1 \cdots \overline u_m$ are upper flow limits. The constraints on tank levels are established by the tank capacities and the necessary water reserve required at all times to handle emergencies effectively. The set $\mathcal{H}$ is defined as
\begin{equation}
% \mathcal{H}=\{[h_1\cdots h_n]^T \in \mathbb{R}^{n} \mid \tilde h_1\leq h_1 \leq \overline h_1, \cdots, \tilde h_n \leq h_n \leq \overline h_n \}
 \mathcal{H}=\{[h_1\cdots h_n]^T \in \mathbb{R}^{n} \mid \forall i : \tilde h_i\leq h_i \leq \overline h_i \}.
 \label{eq:stateConstr}
\end{equation}

 In this work, it is assumed that the pressures in the WDN would be regulated by the elevated tanks and PRVs.  Because of this, pressure constraints are not included in the controller formulation \eqref{eq:MPCForm}. Water losses because of the pressure levels in the network are assumed to be captured by the total aggregated demand $d_{a}$.

 Linear model \eqref{eq:discState} provides only an approximation of the WDN and does not capture its complete representation. Also, it is assumed that the future values of the total demand $d_a$ can only be estimated.  Consequently, the true state trajectories $h$ will differ from the predicted states $h^t$. Although the predicted states $h^t$ adhere to the state constraints by the construction of the controller problem \eqref{eq:MPCForm}, the actual trajectories $h$ may not satisfy these constraints, particularly when the predicted states approach the boundaries of the state constraint set $\mathcal{H}$. Even small deviations of actual trajectories $h$ from the predicted states $h^t$ can result in violations of the state constraints when the predicted states are close to the boundaries. To address this issue, exponential barrier-like functions are used to keep the predicted trajectories away from the boundaries of state constraints so that actual trajectories also satisfy the state constraints. First, state constraints \eqref{eq:stateConstr} are rewritten as
\begin{equation}
F_i(h) \leq 0, \quad i=1,2,\cdots 2n
\label{eq:stateIneq}
\end{equation}
where $F_{2i}(h)=\tilde h_{i}-h_{i}, ~ i=1,\cdots n$ and $F_{2i-1}(h)= h_{i}-\overline h_{i}, ~ i=1,\cdots n$. The cost function terms are then defined as
\begin{equation}
J_{h_i}(h)=e^{a_i(F_i(h)+b_i)} \quad i=1,\cdots, 2n
\label{eq:barrCost}
\end{equation}
 where $a_i,b_i \in \mathbb R_{> 0}$. The parameters $a_i$ and $b_i$ specify a region next to the boundaries of the state constraint set $\mathcal{H}$, wherein the cost function $J_{hi}$ gets high values. When an optimizer solves the problem \eqref{eq:MPCForm}, it automatically finds solutions $h^t$ that don't go into this expensive region. As long as the predicted states $h^t$ stay away from the edges of $\mathcal{H}$, the actual states $h(t)$ will also stay within those limits, provided that the deviation of the actual states from the predicted remains sufficiently small. In simple terms, the cost function terms $J_{h_i}$ penalize being too close to the edges of $\mathcal{H}$. Therefore, predicted states $h^t$ stay away from the edges, ensuring that the actual paths $h(t)$ stay within $\mathcal{H}$.

The cost function $J(h_j^t,u_j^t,t,P_{pv},c)$ incorporates the expenses associated with purchasing electricity from the electrical grid. The primary energy source for the WDN is obtained from PV panels, and in cases where the power generated by the PV panels is less than the power required by the WDN, additional electricity is purchased from the grid to meet the demand.  When PV power is more than the power required by the WDN, no electricity is bought from the grid. Therefore, the electricity purchased from the grid is represented as
\begin{equation}
\label{eq:pvgrid}
P_{grid}(t)=\max(0,P_{p}(t)-P_{pv}(t))
\end{equation}
 where $P_{grid}$ is the power purchased from the grid, $P_{p}$ is the total power used by the pumps and $P_{pv}(t)=X_{\eta}^i$ is the power generated by the PV panels, where $\eta$ and $i$  refer to the specific day and time index within that day that aligns with the moment $t$. Note that the right-hand-side of \eqref{eq:pvgrid} is equal to the rectified linear unit (RELU) function of $P_{p}-P_{pv}$. As the grid power $P_{grid}$ is a non-differentiable function, RELU, of $P_{p}-P_{pv}$, it is approximated with the softplus function 
 \begin{equation}
sp(P_{p}(t)-P_{pv}(t))=\frac{1}{\beta}\log(1+\exp{\beta( P_{p}(t)-P_{pv}(t)}))
\end{equation}
where $\beta$ is a constant. This function is widely used as a smooth approximator of the RELU function.
The power consumed by the pumping station $i$ is equal to ${u}_{i}(p^{out}_i-p^{in}_i)/\lambda_i$, where ${u}_{i}$ is the flow out of the pumping station, $p^{out}_i$ and $p^{in}_i$ are the outlet and inlet pressures of the station  measured in $N/m^2$ and $\lambda_i$ is the efficiency.  It is assumed that each pumping station has several pumps of the same size connected in parallel. The number of active pumps is controlled by a low-level controller to minimize power consumption. Then, the efficiency of the pumping stations remains relatively stable, allowing the assumption of constant efficiency, $\lambda_i=\lambda, \: \forall i \in \{1,2\cdots m\}$ \cite{KALLESOE201124}.  The outlet pressures $p^{out}\in \mathbb{R}^{m}$ are given as the output of the linear model
\begin{equation}
p^{out}(t)=Ch(t)+Du(t)
\label{eq:heads}
\end{equation}
where $C  \in \mathbb{R}^{m\times n}$ and $D \in \mathbb{R}^{m\times m}$ are found using system identification on data generated by the EPANET model.  { The inlet pressures $p^{in} \in \mathbb{R}^{m}$ refer to the pressures of distinct water reservoirs where pumps draw water. It is assumed that these pressures remain constant, implying that the water levels in the associated reservoirs are considered to be unchanging. While daily variations in reservoir water levels are possible, this assumption is made due to the lack of available data, as the reservoir levels on EPANET models are always constant.  However, in real situations, inlet pressures can be modeled more accurately like outlet pressures are modeled in \eqref{eq:heads} to capture daily variations if the measurements for water reservoir levels are available.} The power used by the pumps at time $t$ is then equal to $P_{p}(t)=u(t)^T(p^{out}(t)-p^{in}(t))/\lambda$. The overall cost function includes both the electricity expense term and the constraint barrier functions   
\begin{multline}
J(h(t),u(t),t,  P_{pv},c)=\sum_{i=1}^{2n}J_{h_i}(h(t))+\\ \sum_{k=0}^{\frac{\Delta_t}{\Delta_{pv}}-1 }c(t)sp(P_{p}(t)-  P_{pv}(t+k\Delta_{pv})).
\label{eq:CostFunc}
\end{multline}
It is assumed that the sampling time for PV data, $\Delta_{pv}$, is shorter than the system sampling time, $\Delta_t$, resulting in multiple PV power values being available within the time period from $t$ to $t+\Delta_t$. The electrical costs are then evaluated individually for each distinct PV power value $ P_{pv}(t+i\Delta_{pv})$ in the interval.

The control problem given in equation \eqref{eq:MPCForm} is solved with a scenario-based approach in which $S \in \mathbb{N}$ number of scenarios are generated. The only stochastic variable appearing in the cost function $J(h(t),u(t),t,  P_{pv},c)$ is the PV power $ P_{pv}$. The PV power values $ P_{pv}$ are sampled following the procedure outlined in Section \ref{sec:PvSample}. However, there is a difference in the approach for sampling the $\zeta_{\delta_{\eta}}$ values which are obtained from distributions \eqref{eq:updErr} before sunrise and \eqref{eq:updErr2} after sunrise. In Section \ref{sec:PvSample}, a rejection sampling technique is used so that \eqref{eq:sampleRule} is satisfied. Here, the rejection sampling technique is not used, and all sets of $\zeta_{\delta_{\eta}}$ generated from independent normal distributions are accepted. This adjustment is made to prioritize faster sampling, even though it results in a decrease in the level of approximation to the actual probability distribution. Consequently, the samples produced using this method may not be as precise as those generated following the approach outlined in Section \ref{sec:PvSample}. This discrepancy leads to a less accurate approximation of the cost function defined in \eqref{eq:MPCForm}. As a result, the pump schedules found through approximating the solution of \eqref{eq:MPCForm} with this sampling method are suboptimal, implying diminished cost-saving compared to the case where sampling is done as explained in Section \ref{sec:PvSample}. {If this control method were to be implemented for a real WDN without speed concerns, the rejection sampling could still be used. However, in this paper, the simulations for the optimization of PV values take a significant amount of time (see Section \ref{sec:Pvopt}). Therefore, to accelerate the simulations, the rejection sampling was not used in this paper.}

The daily water demand profiles are closely similar with only small variations. This is used to increase the chance of finding a solution to the optimization problem \eqref{eq:MPCForm}. The idea revolves around maintaining the tank levels at the start of each day within a specific range. If a pumping schedule can steer the system's state trajectories close to their initial values by the end of a day, it is probable that a similar schedule could be applied on the next day since the tank levels at the beginning of each day are close to one another. Therefore, the horizon $N(t)$ is chosen to target the beginning of each day and expressed as
\begin{equation}
N(t)=(T_{day}-t\bmod T_{day})/\Delta_t
\end{equation}
where $T_{day}$ is the duration of a whole day. To define the terminal set $\mathcal{{H}}_{tf}$ in controller problem \eqref{eq:MPCForm} that the trajectories should return to at the start of each day, the optimal periodic trajectory of the system should be defined first. The optimal periodic trajectory is defined just as in \cite{BROOMHEAD201530} as the solution of
\begin{subequations}
\label{eq:optTraj}
\begin{align}
 &(\mathbf{u^*},\mathbf{h^*})=\argmin_{{u_i},{h_i}}  \sum_{i=0}^{(T_{day}/\Delta_t)-1} J({h_i},{u_i},t,P_{pv}^*,c^*)
\\
 &{h_i}=A_d{h_{i-1}}+B_{d1}{u_{i-1}}+B_{d2} d_{a}^*(i-1)
\\
 &{u_i}  \in \mathcal{U} \subseteq \mathbb{R}^{m}&
 \\
 &{h_i}  \in \mathcal{H \ominus \mathcal{W}} \subseteq \mathbb{R}^{n}&
 \\
 &{h_0}=h_{T_{day}/\Delta_t}
  \label{eq:period}
\end{align}
\end{subequations}
where $d_{a}^*(i) \; i=0, 1\cdots (T_{day}/\Delta_t)-1$ is the average daily demand profile obtained from the past measurements, $P_{pv}^*$ is the average daily PV profile, $c^*$ is the average daily electricity price profile, and the set $\mathcal{W}$ is a design choice. The resulting state trajectory $\mathbf{h^*}=[h^*_0 \cdots h^*_{T_{day}/\Delta_t}] \in 
\mathbb{R}^{n\times(T_{day}/\Delta_t+1)}$ is the optimal periodic trajectory as dictated by the periodicity constraint \eqref{eq:period}. The terminal set for the controller problem \eqref{eq:MPCForm} is chosen as the points close to the $h_{T_{day}/\Delta_t}^*$. This decision is based on the idea that if the actual tank levels would revolve around the optimal periodic trajectory $\mathbf{h^*}$, the associated economic costs would remain low. By driving tank levels close to $h_{T_{day}/\Delta_t}^*$ at the end of each day, the actual tank levels would remain close the optimal periodic trajectory $\mathbf{h^*}$, whenever the power, electricity price and demand profiles of a day is close to the average profiles $P_{pv}^*,c^*,d_{a}^*$. After finding the solution to \eqref{eq:optTraj}, the terminal state constraint set $\mathcal{H}_{tf}$ for the controller problem \eqref{eq:MPCForm} then can be written as
\begin{equation}
\label{eq:terminal}
 \mathcal{{H}}_{tf}= \mathcal{B}_r( h_{T_{day}/\Delta_t}^*)
\end{equation}
where $\mathcal{B}_r(h_{T_{day}/\Delta_t}^*)$ is the open ball centered at $h_{T_{day}/\Delta_t}^*$ with radius $r$.
 If the problem \eqref{eq:MPCForm} becomes infeasible at any time step $t$, the second term of the input sequence from the previous step $u_1^{t-\Delta_t}$ is applied to the network. The reason behind using the narrower set $\mathcal{H \ominus \mathcal{W}}$ as the state constraint set, rather than $\mathcal{H}$, when computing the optimal periodic trajectory as defined in equation \eqref{eq:optTraj}, is to prevent the violation of state constraints by the state trajectories $h(t)$ when the problem \eqref{eq:MPCForm} becomes infeasible. Normally, if the optimization problem is feasible,  the state trajectories $h(t)$ are expected to stay close to the optimal periodic trajectory $\mathbf{h^*}$ because of the terminal state constraint $\mathcal{H}_{tf}$. However, the state trajectories $h(t)$ might deviate more from the optimal periodic trajectory $\mathbf{h^*}$ when the optimization problem \eqref{eq:MPCForm} becomes infeasible. To mitigate this, by confining the optimal periodic trajectory within the narrower set $\mathcal{H \ominus \mathcal{W}}$ instead of $\mathcal{H}$, it becomes more likely that the state trajectories $h(t)$ will remain within the bounds of the state constraint set $\mathcal{H}$ even when the optimization problem is infeasible. The design choice for set $\mathcal{W}$ involves a trade-off between feasibility and cost-effectiveness. If it is chosen to be big, the optimal periodic trajectory $\mathbf{h^*}$ will be confined to the more inner parts of $\mathcal{H}$ which will lead to state trajectories $h(t)$ revolving more in the inner side of $\mathcal{H}$, resulting in a less cost-effective but safer setting in terms of satisfaction state constraints. Conversely, a smaller $\mathcal{W}$ has the opposite effect, less conservative and more cost-effective.
 
 In this work, the error set $\mathcal{W}$ is chosen as the set containing the possible one-step errors between the linear model \eqref{eq:discState} and the actual EPANET model of WDN. This set is found by investigating the dataset generated by the EPANET model and taking the difference between the tank levels generated by EPANET and the linear model (see Section \ref{subsection:Sim}).

 \section{Optimizing PV Installation Amount}
 \label{sec:Pvopt}
 This section presents a  new method to optimize the amount of PV installation for a WDN. The objective is to find the amount of PV that minimizes the overall expected cost of the WDN throughout the lifespan of the PV panels, considering both operational expenses (OPEX) and capital expenditures (CAPEX).  OPEX refers to the costs associated with maintaining the PV panels and purchasing energy from the grid to power the pumps, while CAPEX encompasses the installation costs of the PV panels. The problem is written as
 \begin{align}
 \label{eq:PvOpt}
     \min_{x}   \E[J_{c}(x)+\sum_{i=1}^{\ell_{pv}}J_{o}(x,P^i_{grid})]
 \end{align}
where $x$ is the PV amount measured by its initial power production under normal conditions $x=P^1_{STC}$ with $P^i_{STC}$ representing the $i^{th}$ year's power production under normal conditions, $P^i_{grid} $ representing the power bought from the grid for the year $i$,  $J_{c}(x)$ is the CAPEX, $J_o(x,P^i_{grid})$ is the yearly OPEX, and $\ell_{pv}$ is the lifespan of the PVs. The expected cost is determined under some assumptions. First, it assumes that the climate will remain constant throughout the lifespan of the PV panels. Consequently, the PV power model, developed using historical weather data, provides an accurate representation of future power production. Although this assumption is not entirely accurate, the potential decrease in PV power production is estimated to be small in the lifespan of a PV panel, at most $4\%$ in 25 years according to Figure 3 in \cite{climate}. Second, it assumes that the water demand profile will remain unchanged throughout the lifespan of the PV system. This assumption is also not entirely true as the population in the area supplied by the pumps might change because of various reasons. Let's consider a situation where a facility determines an approximate optimal amount of PVs with this assumption and installs them. If the population in the serviced area grows, a reasonable expectation would be an increased optimal number of PVs. In this case, the initial installation of PVs doesn't hinder achieving the new optimal amount, as the WDN facility can add more PVs to match the increased demand. Conversely, if the population decreases for unknown reasons, the anticipated optimal PV amount for the reduced population would be smaller than the initial installation. In such a case, it could be argued that the initially installed PVs were excessive. However, due to the difficulty in predicting these demographic shifts, the assumption of constant water demand is retained. Additionally, the linear model \eqref{eq:discState} is assumed to accurately represent the WDN throughout the lifespan of PVs. This means that factors such as the aging of WDN elements, including pipes, are not considered in this work. This assumption is made because of the challenge of anticipating alterations in the WDN and their potential impact on the parameters of the linear model. %This is an assumption that is hard to see through. However, in the case of an increase in the water demand, the WDN facility can always install more pumping stations. 

Various factors, such as the panel type, market competition within the PV industry, the purchasing country, and the scale of the installation can influence the pricing of PV panel installation. Typically, as the quantity of PV panels increases, the cost per kilowatt (kW) tends to decrease. While this paper adopts a constant unit cost based on an average price, a more detailed model could be used by engaging in discussions with a specialized PV installation company to consider tailored unit costs. Using a more detailed cost model for installation could lead to more savings in the total costs of the WDN. This is because, usually, as you increase the number of PV panels, the cost per panel goes down, thanks to discounts from buying in bulk. The installation cost can be written as
\begin{align}
    J_{c}(x)=a_{ins}x \label{eq:PVinst}
\end{align}
where $a_{ins}\in \mathbb{R}$ is the constant unit cost. OPEX covers the maintenance costs and the costs of purchasing power from the grid and can be written as
\begin{align}
    \label{eq:OPEX}
    J_{o}(x,P^i_{grid})=a_mx+
    \sum_{j=1}^{365}\sum_{l=1}^{T_{day}/\Delta_{pv}}c(t)P^i_{grid}(t)\Delta_{pv}
\end{align}
where $a_m\in \mathbb{R}$ is the unit maintenance cost, $t=jT_{day}+l\Delta_{pv}$ is the time of the year, $P^i_{grid}(t)$ is the power bought from the grid at time $t$ in year $i$. The power bought from grid $P^i_{grid}(t)$ depends deterministicly on the yearly PV power samples $P^i_{PV}(t)$. The statistical properties of $P^i_{PV}(t)$ vary with the year index $i$ due to changes in the standard test condition power $P^i_{STC}$ over the years. The standard test condition power $P^i_{STC}=\lambda^i_{pv}x, ~ i=1,\cdots \ell_{pv}$ changes over time because the efficiency of the solar panels, $\lambda^i_{pv}$, decreases as they age. To simplify the analysis, a constant efficiency model for the PVs is used. This ensures that the standard test condition power $P_{STC}=\lambda_{pv}x$ remains constant throughout the lifespan, with $\lambda_{pv} \in \mathbb{R}$ representing the constant efficiency whose value is set to the average efficiency over the lifespan of PVs $\lambda_{pv}=\frac{1}{\ell_{pv}}\sum_{i=1}^{\ell_{pv}}\lambda^i_{pv}$. By assuming this, the parameters of the PV power model can be kept constant over the years, ensuring that the statistical properties of the yearly solar power production $P^i_{PV}(t)$ and the power bought from the grid $P^i_{grid}(t)$ remain the same over the years. Therefore, the expected values $\E[J_{o}(x,P^{i_1}_{grid})]$ and $\E[J_{o}(x,P^{i_2}_{grid})]$ are equal for any years $i_1,i_2 \in \mathbb{N}$. Now, the dependency of the OPEX on the year index can be removed by defining $J_{o}(x)=J_{o}(x,P^{i_1}_{grid})$ for some fixed year $i_1 \in \mathbb{N}$. This simplification enables reformulation of the problem \eqref{eq:PvOpt} as 
\begin{align}
    \label{eq:PvOpt2}
    \min_{x}   \E[J_{c}(x)+\ell_{pv}J_{o}(x)]
\end{align}
The reasoning behind these assumptions and the removal of the OPEX dependency on the year index is to avoid extensive computation time required to evaluate the term $\sum_{i=1}^{\ell_{pv}}J_{o}(x,P^{i}_{grid})$ in \eqref{eq:PvOpt} for a given value of $x$. Such calculations would significantly prolong the optimization procedure. We have tried using neural networks to approximate the behavior of the predictive controller, which would have expedited the cost calculation. This approach would have minimized the time needed to simulate the network over the lifespan of PVs and calculate the term $\sum_{i=1}^{\ell_{pv}}J_{o}(x,P^{i}_{grid})$. However, the accuracy of the neural network approximations of the controller was insufficient. As a result, it is decided to maintain the assumption of a constant efficiency $\lambda_{pv}$.

 To approximate the total cost $\E[J_{c}(x)+\ell_{pv}J_{o}(x)]$ for a given amount of PVs $x$, first step is generating the synthetic historical PV data for the given value $P_{STC}=\lambda_{pv}x$ and using the physical production model \eqref{eq:physPV}. { Note that the same historical weather measurements are used for each $x$.} Then, the parameters of the time series representations \eqref{eq:opmultmodel}, \eqref{eq:optMultErrMod}, \eqref{eq:multErr} of the PV production model defined by equations  \eqref{eq:pvModel}, \eqref{eq:EWMA}, \eqref{eq:optMultDef}  are fit to the synthetic historical PV data. Then, one year of PV production data is sampled using the PV model.  Subsequently, the system's tank levels are simulated for one year by treating the sampled one-year PV power data as the real PV production data. The pump flow rates are derived from solving the control problem \eqref{eq:MPCForm} at each time step. For this, the terminal condition of the control problem \eqref{eq:MPCForm} is determined through solving \eqref{eq:period}. At each step,  the amount of power purchased from the grid $P_{grid}(t)$ is calculated using equation \eqref{eq:pvgrid} after the pump flow rates are found. For the simulation of WDN and evolving the tank levels, the approximate system model \eqref{eq:discState} is used instead of the actual EPANET model of the network, as the latter is computationally intensive. Despite using the approximate model, it still takes approximately 6 hours to compute the yearly OPEX $J_{o}(x)$ for a given $x$ and the yearly power profile on a computer with Intel i5-1135G7 2.40 GHz processor and 32 GB RAM. Due to time constraints, a single sample of the yearly PV profile is used to approximate $\E[J_{c}(x)+\ell_{pv}J_{o}(x)]$ for a given $x$. While acknowledging the limitation of having only one sample, it is expected to provide a reasonable approximation of the expected value. This is because essentially 365 days are sampled,  potentially covering both days when the generated solar power is more than the seasonal average and days when the generated solar power is less than the seasonal average in a year. A similar argument can be made about the intraday variations in PV power production. Using the multiplicative correction term model \eqref{eq:multErr}, approximately $365\times 12$ hours of intraday variations are generated for a year, providing a reasonable representation of various intraday scenarios.

 The closed form of the function in \eqref{eq:PvOpt2} is not known as the calculation of the cost involves solving the optimization problem \eqref{eq:MPCForm}, which also means gradient information for the cost function is also not available. Therefore, a direct search method is used, specifically the Nelder-Mead method \cite{NM}, to iteratively update PV amounts and find the amount minimizing the cost function. The Nelder-Mead method is particularly suitable for this problem, as it shows significant improvements in the initial steps and requires fewer evaluations of the cost function compared to other methods \cite{neldermeadConv}. This is important for this application as the calculation of the cost function \eqref{eq:PvOpt2} requires the simulation of the model \eqref{eq:discState} for a year, which is extremely time-consuming.

The sampling of one-year PV data is done in a sequential process. Initially, the daily profile of the first day is assigned the value from one year ago,  $Y_{\eta}=Y_{\eta-365}$, where $Y_{\eta-365}$ is obtained from the synthetic power data. Then, the optimal multiplier $p_{\eta}$ and the multiplicative correction values $\delta_{\eta}$ for the current day $\eta$ are sampled using respectively \eqref{eq:opmultmodel}, \eqref{eq:optMultErrMod} and the power data $X_{\eta}$ is obtained.  Following that, the daily power profile is updated $Y_{\eta+1}$ and the process continues to the next day. The sampling of the optimal multiplier $p_{\eta}$ and the multiplicative correction values $\delta_{\eta}$ are done as explained in Section \ref{sec:PvSample}. The overall process of calculating the yearly OPEX $J_{o}(x)$ is given in Algorithm \ref{alg:opex}.

\begin{algorithm}
\caption{Calculation of yearly OPEX $J_{o}(x)$}\label{alg:opex}
\hspace*{2.3mm} \textbf{Input} $x,Y_{1-365}$ \Comment{PV Amount and the  daily Power profile for the first day} \\ 
\hspace*{2.3mm} \textbf{Output} 
$J_{o}(x)$ 
\begin{algorithmic}
\Procedure{}{}
\State $P_{STC} \gets \lambda_{pv}x$  
\State Produce synthetic PV data with historical weather measurements, namely irradiance ($G$), ambient temperature ($T_{amb}$), and wind speed ($V_{wind}$), using the physical model equations \eqref{eq:physPV}, \eqref{eq:physNorm}, 
 \eqref{eq:ModTemp}.
\State Fit the parameters of the time series representations \eqref{eq:opmultmodel}, \eqref{eq:optMultErrMod}, \eqref{eq:multErr} of the PV production model defined by equations  \eqref{eq:pvModel}, \eqref{eq:EWMA}, \eqref{eq:optMultDef} to the synthetic historical PV data.
\State $Y_{1} \gets Y_{1-365}$
\For{\texttt{$k \gets 1$ to $365$}}
    \State Sample $p_k$ using \eqref{eq:opmultmodel}, \eqref{eq:optMultErrMod}
    \State Sample 
    $\delta_{k}^i, \: i=1 \cdots T_{day}/ \Delta_{pv}$ using \eqref{eq:opmultmodel}, \eqref{eq:optMultErrMod}
    \State $X_{k} \gets p_{k}Y_{k}\odot \delta_{k}$
    \State Get $Y_{k+1}$ using \eqref{eq:yUpdate}
\EndFor
\State Get the terminal state set using \eqref{eq:optTraj}.
\For{\texttt{$i \gets 1$ to $365*24$}}
    \State Solve the problem \eqref{eq:MPCForm} and find the input $u(i \Delta_t)$.
    \State Calculate $P_{grid}(i\Delta_t)$ with \eqref{eq:pvgrid}
    \State Simulate the system with \eqref{eq:discState}.
\EndFor
\State Calculate $J_{o}(x)$ using \eqref{eq:OPEX}.
\EndProcedure

\end{algorithmic}
\end{algorithm}

\section{Application}
\label{sec:application}
\begin{figure}
\resizebox{1\columnwidth}{!}{\includegraphics{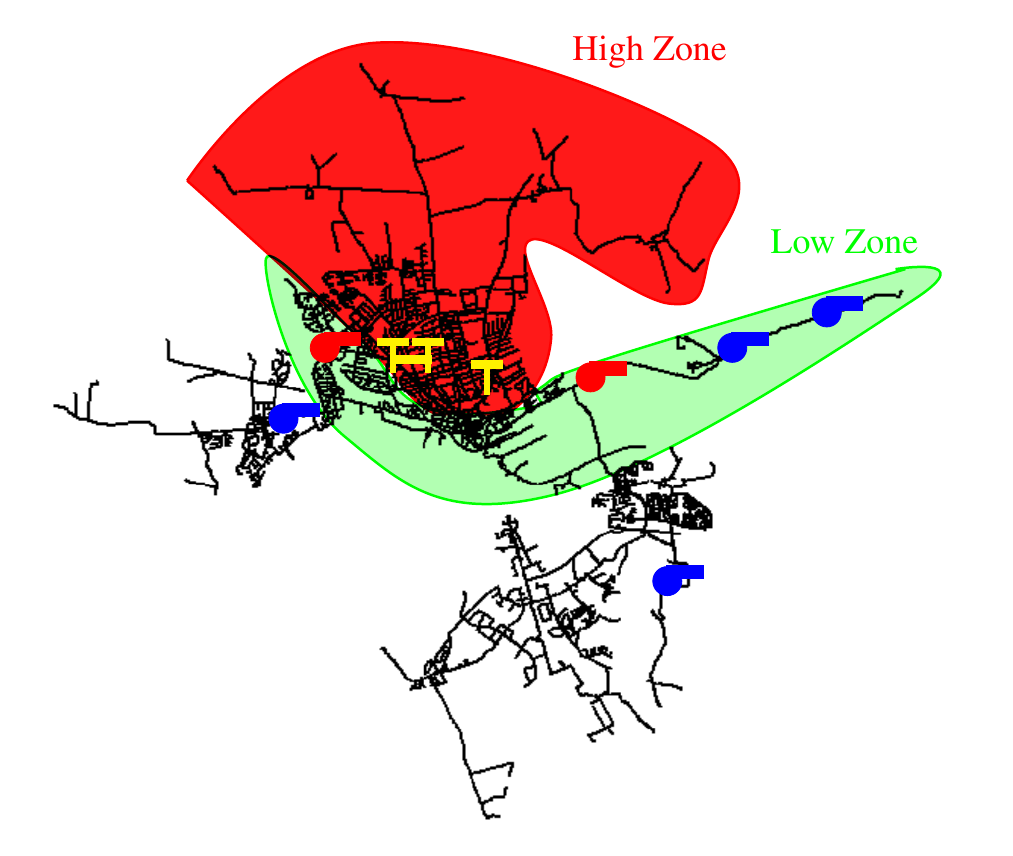}}
\centering
\caption{Water Distribution Network of Randers. The pumping stations to be controlled are shown in red and the remaining in blue. Tanks are shown with a 'T' shaped symbol in yellow.}
\label{fig:network}
\end{figure}

The presented method is used to find an approximation of the optimal installed PV amount for the WDN of Randers, a Danish city. The EPANET model of the WDN, illustrated in Figure \ref{fig:network}, is used to get the relevant information about the network.
The network contains 4549 nodes and 4905 links connecting them and includes 8 pumping stations. Out of these, 6 are shown in the figure while the remaining 2 are located where tanks are placed. Our goal is to control the flow rates of the two pumping stations, shown in red color in Figure \ref{fig:network}, and find an approximation of the optimal PV amount for them, while the others operate according to pre-determined strategies. The decision to control and determine PV amounts for only two stations was made in collaboration with the WDN company due to limited space in other pumping stations for PV installations. Also, there are limited opportunities to optimize energy usage in the other pumping stations since they mainly serve areas without tanks, making significant flow rate adjustments difficult due to pressure requirements. The same method could be used for any number of pumping stations in the network. The only difference would be the identification of the parameters in the linear model \eqref{eq:reducedModel} for the desired pumping stations to be controlled. The stations that will be controlled provide water mostly to the High Zone (HZ) and Low Zone (LZ) shown in Figure \ref{fig:network}. Additionally, there are 3 tanks in the HZ, two of them connected with a pipe and the third standing alone.

  In the linear model \eqref{eq:reducedModel}, the number of inputs is equal to $m=2$ since two pumping stations are being controlled. The number of state variables is equal to the number of tanks in the network. However, two of the tanks are directly connected by a large pipe. This means that whenever there's a difference in their water levels, water quickly moves through the pipe connecting them. As a result, the water levels in both tanks, $h_1$ and $h_2$, are almost always very close $h_1 \approx h_2$. Therefore, we can treat the levels of these two tanks, $h_1$ and $h_2$, as a single state variable ${h}_{1,2}$. {To identify the values of constant matrices $A, B_1, B_2$ in the model \eqref{eq:reducedModel}, first the EPANET model is simulated for 20 days with a 5-minute simulation time step,  generating data for around 6000 time instances. For each day, different values for the initial tank levels are used and flow rates of the two controlled pumping stations are set to values varying from $0$ to maximum flow rate $u_{max}$.} Control laws for the remaining pumping stations are predefined within the EPANET model. Next, the flow of the pipes connected to the tanks is approximated as linear functions of the tank levels $h(t)=[{h}_{1,2}(t)~{h}_{3}(t)] \in \mathbb{R}^{2}$, pump flows $u(t)=[{u}_{1}(t)~ {u}_{2}(t)] \in \mathbb{R}^{2}$ and the aggregated demand $d_a(t)$ using the simulation data. The total demand for the High and Low Zones is used as the aggregated demand in the model since these areas are primarily supplied by the controlled pumps. The linear model of the system \eqref{eq:reducedModel} is obtained by putting the linear approximations of the pipe flows into the tank level equation \eqref{eq:tankLevel}. Note that since other pumping stations were controlled using predefined laws in EPANET, their effect on the evolution of the tank levels is reflected in the data generated by EPANET. Therefore, the network model \eqref{eq:tankLevel} contains information about the effect of other pumping stations as the parameters of the model are fitted using this dataset. Also, the effect of the other pumping stations on the levels of the three tanks in HZ is small either because the capacities of pumping stations are small or they mostly supply water to the other areas.

\subsection{Simulation Results}
\label{subsection:Sim}
This section presents the simulation setup and results for the Randers WDN. The historical weather measurement data is taken from the Open Data API of the Danish Meteorological Institute (DMI). The values of the parameters $k_i, i=1 \cdots 6$ are given in \cite{HULD20113359}, and the values of $\mu_0,\mu_1$ in \cite{BARYKINA2017401} are used for the physical model \eqref{eq:physPV}.  From the data generated by EPANET, we empirically identified the set of possible differences between the one-step tank level change estimation of the linear model \eqref{eq:discState} and the corresponding change in the EPANET model $h_1^t-h(t+\Delta_t)$. This range was determined to be $\interval{-0.1}{0.1}\times \interval{-0.1}{0.1}$. The parameters of the exponential barrier functions $J_{hi}$ were then set to $a_i=80$ and $b_i=0.2$ for all $i$. With these parameter choices, the values of $J_{hi}$ become significantly higher than the other terms in the cost function \eqref{eq:CostFunc} within the region where $F_i(h) \in \interval{-0.1}{0}\times \interval{-0.1}{0}$, as defined in \eqref{eq:stateIneq}. Therefore, as long as the next predicted state $h_1^t$ is not in the region where $J_{hi}$ takes high values, the actual next state $h(t+\Delta_t)$ will satisfy the state constraints $F_i(h)\leq0$. This is because $h_1^t-h(t+\Delta_t)$ is found to be always smaller than $0.1$ in our EPANET data analysis.  In the formulation of optimal periodic trajectory, $\mathcal{W}$ is also defined as the set containing all possible one-step errors $h_1^t-h(t+\Delta_t)$, that is $\mathcal{W}=\interval{-0.1}{0.1}\times \interval{-0.1}{0.1}$. Both decisions regarding $a_i,b_i$ parameters and $\mathcal{W}$ are a matter of design. Opting for a larger $\mathcal{W}$ could be a safer choice in terms of state constraint satisfaction but this might make the control less cost-effective. The same could be said about $a_i,b_i$ parameters and the region defined by them. The maximum tank levels are 3m for $h_1$ and $h_2$, and 2.8m for $h_3$, while the minimum tank level is set to half-full. The maximum pump flow is set to 100 liter/sec and the sampling time $\Delta_t$ to 1 hour, so the control input is recalculated at each hour. In all simulations, the yearly electricity prices $c(t)$ are set to the day-ahead electricity prices in Denmark from the year 2022. The daily total demand of HZ and LZ $d_{a}(t), \; t=0, \Delta_t\cdots T_{day}/-\Delta_t$ is obtained from the EPANET model and the network model \eqref{eq:discState} is simulated using these values in all simulations. Given the assumption of constant daily demand, it is inferred that water losses, encapsulated in the total demand, remain stable throughout the entire lifespan of PVs. We can only use an estimation $d^t_{a}$ of the daily total demand values in our calculations of pump schedules \eqref{eq:MPCForm} since we assume that the total demand can only be estimated. However, historical water demand data of Randers WDN is not available so the total demand $d_a(t)$ cannot be estimated. Instead, a historical demand data set from another facility in Denmark is used. First, the daily average demand of the demands in the dataset is found. Then, a small perturbation signal is obtained by taking the difference between the daily average demand and the demand on a specific day. This small perturbation is added to the actual total demand $d_{a}(t) \; t=0, \Delta_t\cdots T_{day}/-\Delta_t$ of Randers WDN resulting in a perturbed daily demand that imitates the estimated demand $d^t_{a}$. In each simulation, a different day from the data set is used, so the assumed estimated demand is different each time. The calculations are based on the assumption that the PV lifespan is 25 years  $\ell_{pv}=25$ \cite{lifespan}. However, we also report the cost values for different lifespan values. In \cite{7870691}, it is reported that the degradation rate of the PVs in Aalborg, Denmark is around $0.15 \%$. The value of average efficiency $\lambda_{pv}$ is set using this information to its average value over the lifespan of the PVs as $\lambda_{pv}=1-0.0015\ell_{pv}/2$. The unit maintenance cost for PVs is set to $a_m=17 \;   \text{\euro} /kW$ \cite{Wiser2020BenchmarkingUP}. The unit installation price is set to $a_{ins}=2 \;  \text{\euro} /W$, which is decided based on the market prices in Denmark.

\begin{figure}
\centering
\begin{subfigure}{0.27\textwidth}
         \centering
         %\caption{}{\includegraphics[width=\textwidth]{figures/25.pdf}}
         \caption{}{\includegraphics[width=\textwidth]{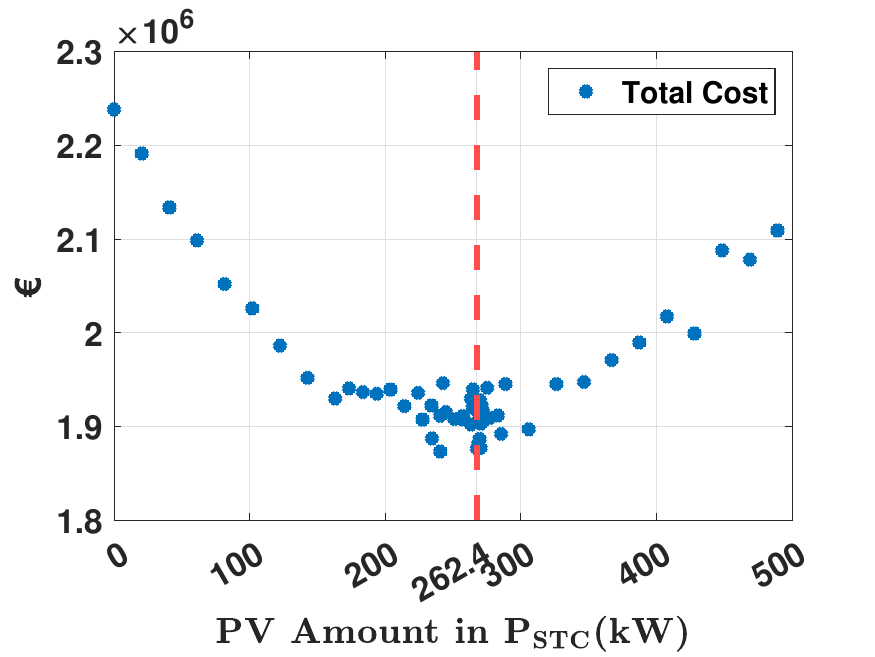}}
         \label{fig:TotalCost}
\end{subfigure} 
\\
\begin{subfigure}{0.27\textwidth}
         \centering
         %\caption{}{\includegraphics[width=\textwidth]{figures/networkCosts.pdf}}
         \caption{}{\includegraphics[width=\textwidth]{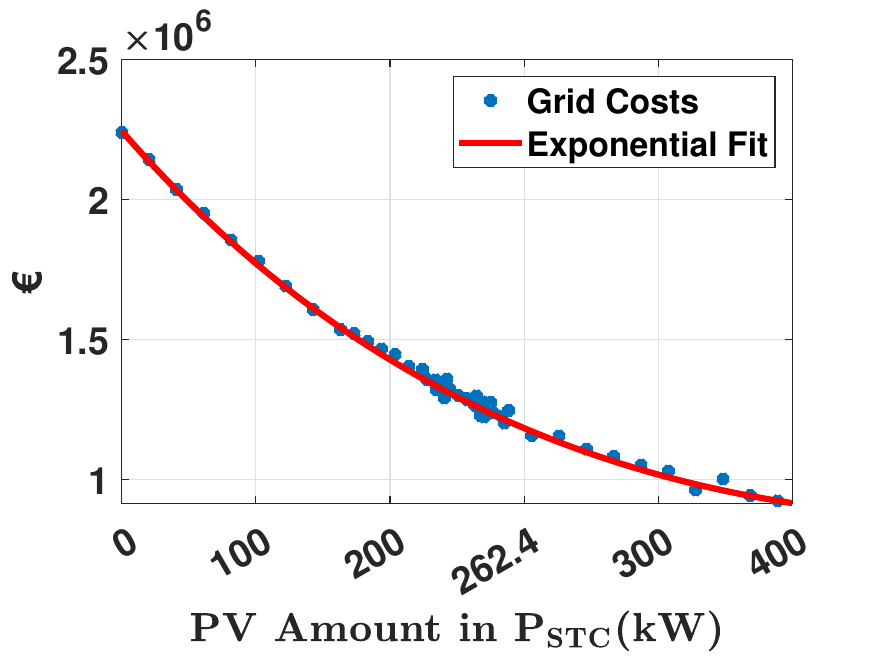}}
         \label{fig:networkElec}
\end{subfigure}
\\
\begin{subfigure}{0.27\textwidth}
         \centering
         \caption{}{\includegraphics[width=\textwidth]{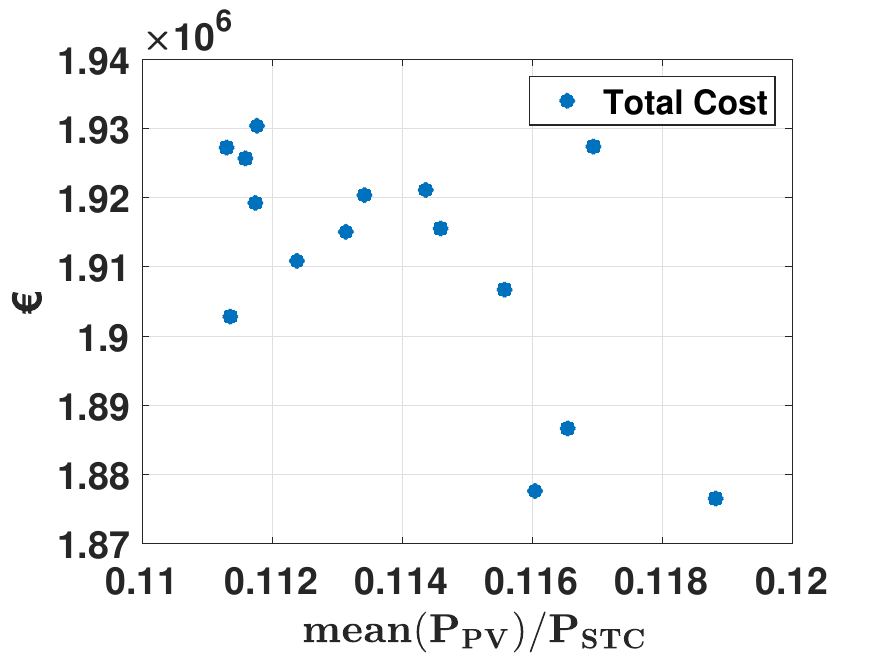}}
         \label{fig:ratio}
\end{subfigure}
\caption{(a) Total cost of the WDN over the lifespan $\ell_{pv}=25$ of the PV panels including the installation and maintenance costs of the PV panels and the electricity costs of the WDN for a given amount of PVs, $J_{c}(x)+\ell_{pv}J_{o}(x)$. (b) The total costs of power bought from the grid and an exponential function fitted to the data. (c) Total cost of the WDN and the ratio of mean PV production of sampled yearly PV profile and the installed PV amount.}
\label{fig:25year}
\end{figure}

The overall cost of the WDN is determined by simulating the linear model \eqref{eq:reducedModel}. The complete calculation process for obtaining the total cost associated with a specific number of PVs typically requires approximately six hours. The calculated cost values for different amounts of PVs encountered throughout the iterations of the Nelder-Mead method are reported in Figure \ref{fig:TotalCost}. By employing the Nelder-Mead Algorithm, an approximation of the optimal PV amount was found to be 262.4 kW. Compared to the costs when no PVs are installed, there is an approximate 14.5 \% decrease in the total costs of the WDN. The total cost exhibits a quadratic trend with noticeable fluctuations, particularly in the vicinity of the approximate optimal value. It's worth noting that the clear trend in cost values supports our expectation that a one-year PV sample should give us a reasonably close estimate of the expected cost. The fluctuations in the cost values stem from the inherent noise introduced by the yearly PV profiles, which are generated through sampling from stochastic variables. These fluctuations around the approximate optimal value are investigated in Figure \ref{fig:ratio}. The figure shows how the total cost of the network is associated with the ratio of mean PV power production from the yearly sampled PV profile to the installed PV amount. Specifically, it focuses on PV amounts that are close to the approximate optimal value. Notably, higher mean PV profiles are associated with relatively lower costs, while lower mean PV profiles result in higher costs. The costs of purchasing power from the grid are given in Figure \ref{fig:networkElec} and approximated with an exponential function fitted to the data. This is the only stochastic term in the total network cost $J_{c}(x)+\ell_{pv}J_{o}(x)$ since installation and maintenance costs are deterministic functions of the installed PV amount. {Assuming the fitted exponential function represents the expected electricity costs, meaning  $J_o(x)$ is considered deterministic and given by the summation of PV installation costs and the exponential fitting of the electricity costs, the optimal PV amount is estimated to be 246.1 kW.} Now, there are two values: one returned by the Nelder-Mead method and another from the exponential fitting of electricity costs. Nelder-Mead method quickly gets close to the vicinity of optimal value but it does not converge to the optimal value for stochastic optimization problems. Therefore, the value obtained through the Nelder-Mead method can only be considered to be in the vicinity of the optimal value. On the other hand, the value obtained by fitting an exponential function to the grid costs, as shown in Figure \ref{fig:networkElec}, may be deemed more reliable, given the evident exponential trend in the grid costs.  Nonetheless, for enhanced precision in results, additional evaluations of total WDN costs under varying amounts of PVs should be conducted.

\begin{figure}
\centering
\begin{subfigure}{0.27\textwidth}
         \centering
         %\caption{}{\includegraphics[width=\textwidth]{figures/25wo.pdf}}
         \caption{}{\includegraphics[width=\textwidth]{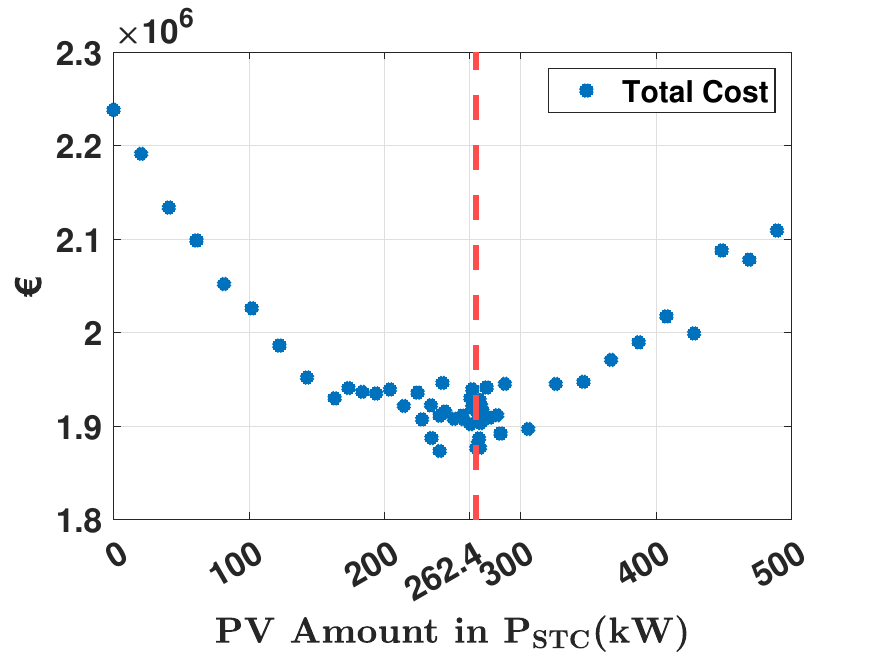}}
\end{subfigure} 
\\
\begin{subfigure}{0.27\textwidth}
         \centering
         %\caption{}{\includegraphics[width=\textwidth]{figures/30.pdf}}
         \caption{}{\includegraphics[width=\textwidth]{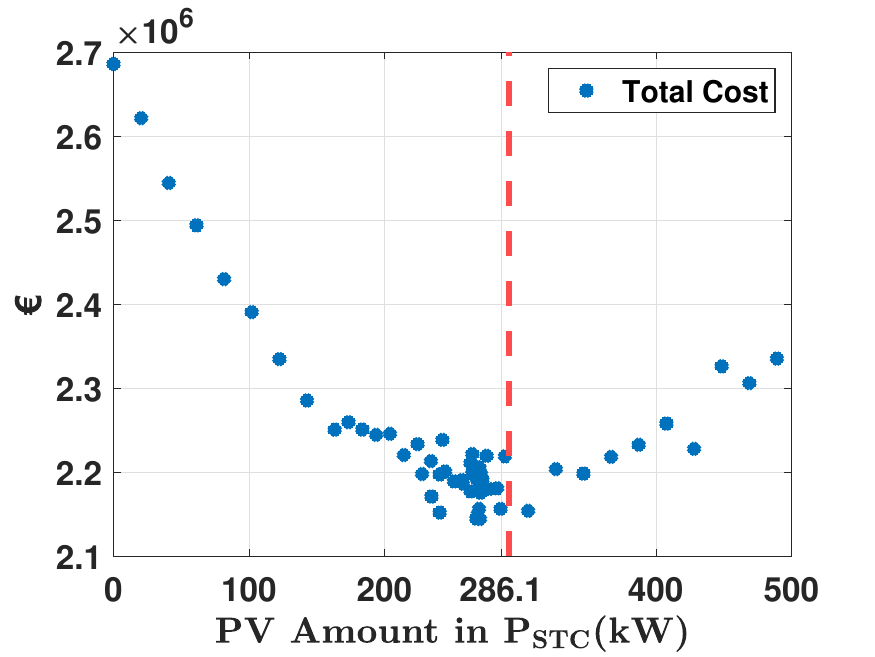}}
\end{subfigure} \\
\begin{subfigure}{0.27\textwidth}
         \centering
         %\caption{}{\includegraphics[width=\textwidth]{figures/35.pdf}}
         \caption{}{\includegraphics[width=\textwidth]{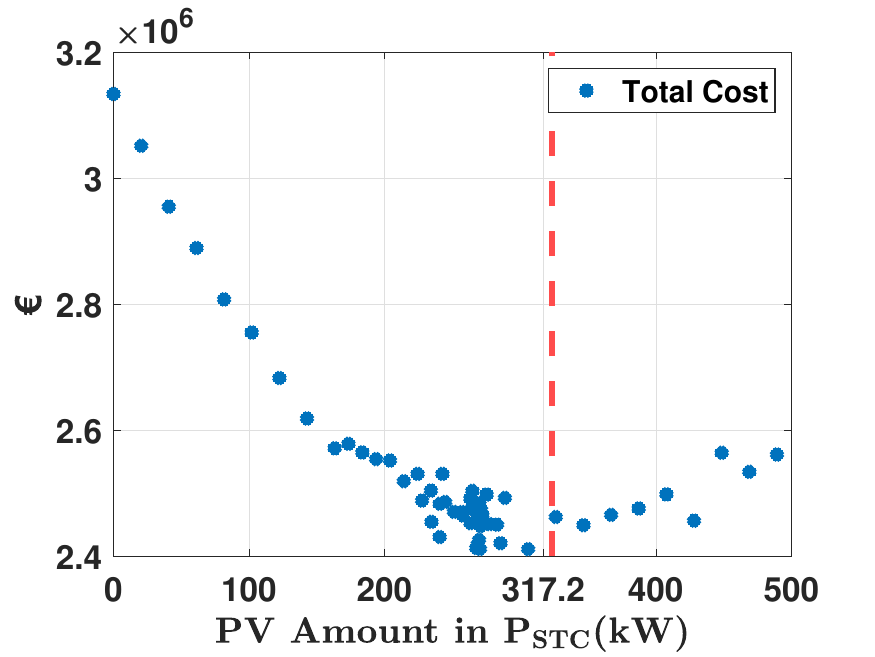}}
\end{subfigure}
\caption{Total costs for different lifespans of PVs $\ell_{pv}=25,\ell_{pv}=30,\ell_{pv}=35$. The red dashed lines represent the approximated optimal quantity of PV panels, obtained by fitting an exponential function to network costs and minimizing the expected total cost based on this exponential function.}
\label{fig:LifeSpan}
\end{figure}

The costs for different life spans $\ell_{pv}=25,\ell_{pv}=30,\ell_{pv}=35$ are given in Figure \ref{fig:LifeSpan}. As the life span increases, we anticipate that the optimal PV amount will exceed the value observed in the $\ell_{pv}=25$ case. This expectation arises from the fact that with a longer life span, OPEX $J_{o}(x)$ increases at a slower rate for higher PV quantities, as they can generate more free power for the WDN. On the other hand, CAPEX $J_{c}(x)$ remains constant. This phenomenon can be observed in the figures, where the cost for larger PV amounts approaches the cost of the approximate optimal amount (262.4 kW) for $\ell_{pv}=25$ as the life span increases. Consequently, the approximate optimal PV amount shifts towards higher values for longer life spans. If we again use the exponential function fitted to the network costs and use it as the expected network cost, the approximate optimal PV amount for $\ell_{pv}=30$ increases to 286.1 kW, and for $\ell_{pv}=35$, it further rises to 317.2 kW. Under this assumption, the cost reduction achieved with $\ell_{pv}=30$ is 18.1\%, while for $\ell_{pv}=35$, it reaches 21.4\%.

It is important to note that the calculation of the approximate optimal PV amount is based on certain assumptions. We used electricity prices from 2022 and assumed a constant water demand throughout the lifespan of the PVs. Additionally, we did not consider the potential impact of climate change on PV power production and water demand. However, these assumptions were applied consistently across all PV amounts, enabling us to identify the approximate optimal PV amount and achieve cost reduction with these assumptions. Moving forward, more accurate representations of water demand and electricity prices are needed to obtain more reliable results. Additionally, a constant PV efficiency model $P_{STC}=\lambda_{pv}x$ is used to ease the computations. Without this simplification, the results would have shown slightly less overall savings. That is because of the convex-like shape of the grid costs (see Figure \ref{fig:networkElec}.). To illustrate this, consider the value returned by the Nelder-Mead method, 262.4 kW. Now assume that PV efficiency is not constant but $P_{STC}=270$ kW for the first half of the PV lifespan and $P_{STC}=250$ kW for the second half. Looking at Figure \ref{fig:networkElec}, it can be seen that the total cost in this situation would be more than if we stuck with a constant $P_{STC}=262.4$ kW throughout. Furthermore, a constant cost for installing PVs is used based on the market price for PV panels. However, in reality, this likely makes the overall cost estimate higher than it would be. Usually, when more PV panels are added, the cost per panel goes down, but the model used in this paper doesn't consider that, so it might have overestimated the total costs. A one-year PV sample is used to simulate the WDN and approximate the total cost \eqref{eq:PvOpt2} for computational purposes. Despite the potential for some inaccuracies, examining Figure \ref{fig:networkElec}, a clear trend in grid costs can be seen. Also, the differences in costs for similar PV quantities are quite small. Therefore, it can be said that using one-year samples didn't lead to significant inaccuracies. { Another factor affecting the results is the half-full requirements on the tanks. With a lower bound on tank levels, pump schedules could have been more flexible, making the WDN operation more cost-effective. This would have resulted in a lower total cost for each given PV amount. However, the impact of this on the optimal PV quantity is not straightforward, as it is challenging to estimate the precise cost reduction for each PV amount.}

\section{Conclusion}
\label{sec:conc}
We have introduced a method to find the optimal amount of PVs to install for a WDN along with a probabilistic PV power production model. The results show that by installing the optimal number of PV systems in the Randers WDN, it is possible to achieve a reduction of 14.5\% in network costs. Note that this number is obtained by assuming the pumps of the WDN will be scheduled with the presented stochastic controller throughout the lifespan of PVs. However, an alternative control method could easily be integrated into the procedure presented in this paper by simulating the WDN with the desired control method. We might question whether assuming a consistent pump scheduling method throughout the lifespan of PVs is practical. This leads to concerns about the reliability of the determined PV amounts if, for example, the WDN facility plans to use different pump scheduling methods. However, as long as the facility adopts a pump scheduling method that minimizes the economic costs of pumps, the resulting pump schedules should not significantly differ from those of other control methods. This is because the primary drivers of cost minimization are the available PV power and grid electricity prices, both of which remain independent of the specific pump scheduling method. Consequently, the proposed PV amounts can be relied upon, provided the WDN facility uses similar pump scheduling methods aimed at minimizing electricity costs for the pumps.  Although the presented procedure provides a detailed approach to determining the required PV installation, it still requires additional steps such as fitting the linear model of the network and simulating the WDN which is time-consuming. It would be desirable to have a simpler method for determining the number of PVs to be installed. However, this is not feasible by solely analyzing the simulation results of a single WDN, as the optimal PV amount depends on various factors, including elevated tank and water pump capacities in the network, average daytime, and even the seasonal variation of daytime throughout the year. To obtain a simplified formula for PV installation based on these parameters, the presented procedure could be applied to different WDNs with varying characteristics in terms of these variables. Then, a regression model could be fit to optimal PV amount as a function of certain WDN properties, climate properties of WDN location etc. This would enable estimating the optimal amount of PVs without having run the WDN simulations. However, to make this work, we need a fast way to calculate or estimate the total costs of the WDN as stated in equation \eqref{eq:PvOpt}.  One way to do this is by using a neural network to mimic the input-output characteristic of the pump scheduling problem \eqref{eq:MPCForm} as running a neural network is a lot faster than solving the given optimization problem. Our initial attempts at training a neural network yielded suboptimal approximations of the pump scheduling problem's input-output characteristics. Further exploration in this direction could bring other potential advantages as well. If successful, a well-trained neural network could accelerate simulations, offering potential benefits such as estimating the total cost of the network \eqref{eq:PvOpt}, without relying on the assumption of constant PV efficiency. This, in turn, could lead to more precise approximations of the overall cost function. As another future work, the linear network model \eqref{eq:reducedModel} could also be improved with the help of demand allocation methods that estimate the demands of each node. If the demand information of each node can be integrated into the linear model \eqref{eq:reducedModel} without making the model complex, this could improve the accuracy of the model without increasing the computational complexity of the controller calculation.


\section*{ACKNOWLEDGMENT}
 We acknowledge Verdo company, Peter Nordahn, and Steffen Schmidt for providing us with the EPANET model and the network information.

\bibliographystyle{IEEEtran}
\bibliography{references}
\end{document}